\documentclass[11pt,a4paper]{article}
\pdfoutput=1
\usepackage{jheppub}
\usepackage{amsfonts, amssymb, amsmath}
\usepackage{mathrsfs}
\usepackage[utf8]{inputenc}
\usepackage[russian,english]{babel}

\usepackage{color}
\definecolor{dark-gray}{gray}{0.20}
\definecolor{gray}{gray}{0.30}
\definecolor{light-gray}{gray}{0.80}
\definecolor{dark-red}{rgb}{0.7,0,0}
\definecolor{dark-green}{rgb}{0.1,0.4,0}
\definecolor{dark-blue}{rgb}{0.3,0.3,0.7}
\definecolor{light-blue}{rgb}{0.8,0.8,1}
\definecolor{blue}{rgb}{0,0,1}
\definecolor{red}{rgb}{1,0,0}
\definecolor{green}{rgb}{0,1,0}

\usepackage{hyperref}
\hypersetup{
	colorlinks=true,
	linkcolor=dark-blue,
	citecolor=dark-red,
	urlcolor=dark-blue,
	linktoc=page
}

\def\cN{{\cal N}}

\def\SO{{\rm SO}}
\def\U{{\rm U}}
\def\SU{{\rm SU}}

\newcommand{\be}{\begin{equation}}
\newcommand{\ee}{\end{equation}}
\newcommand{\bea}{\begin{eqnarray}}
\newcommand{\eea}{\end{eqnarray}}

\title{Factorization of log-corrections in AdS$_4$/CFT$_3$\\ from supergravity localization}

\author[a]{Kiril Hristov}
\author[b]{and Valentin Reys}

\affiliation[a]{Faculty of Physics, Sofia University, J. Bourchier Blvd. 5, 1164 Sofia, Bulgaria}
\affiliation[a]{INRNE, Bulgarian Academy of Sciences, Tsarigradsko Chaussee 72, 1784 Sofia, Bulgaria}
\affiliation[b]{Instituut voor Theoretische Fysica, KU Leuven, Celestijnenlaan 200D, B-3001 Leuven, Belgium}

\emailAdd{khristov@phys.uni-sofia.bg}
\emailAdd{valentin.reys@kuleuven.be}

\abstract{
\noindent We use the Atiyah-Singer index theorem to derive the general form of the one-loop corrections to observables in asymptotically anti-de Sitter (AdS$_4$) supersymmetric backgrounds of abelian gauged supergravity. Using the method of supergravity localization combined with the factorization of the supergravity action on fixed points (NUTs) and fixed two-manifolds (Bolts) we show that an analogous factorization takes place for the one-loop determinants of supergravity fields. This allows us to propose a general fixed-point formula for the logarithmic corrections to a large class of supersymmetric partition functions in the large~$N$ expansion of a given 3d dual theory. The corrections are uniquely fixed by some simple topological data pertaining to a particular background in the form of its regularized Euler characteristic $\chi$, together with a single dynamical coefficient that counts the underlying degrees of freedom of the theory.
}
\date{\today}
\begin{document}
\maketitle

%%%%%%%%%%%%%%%%%%%%%%%%%%%%%%%%%%%%%

%%%%%%%%%%%%%%%%%%%%%%%%%%%%%%%%%%%%%
\section{Introduction and main result}
\label{sec:intro}
%%%%%%%%%%%%%%%%%%%%%%%%%%%%%%%%%%%%%

The development of supersymmetric localization~\cite{Pestun:2007rz} has lead to major progress in the field of holography and the AdS/CFT correspondence in the past decade. There now exist remarkable calculations of observables in different holographic superconformal field theories (SCFTs) on various supersymmetric backgrounds, often beyond the leading large~$N$ approximation, see~\cite{Pestun:2016zxk} and many references therein and thereof. Much of this progress further relies on the closely related idea that the partition functions often localize not only in field space, but also on specific points of the particular supersymmetric background, giving rise to a factorization in terms of elementary building blocks~\cite{Nekrasov:2002qd,Nekrasov:2003vi}. Here we will mostly be interested in 3d SCFTs that were shown to admit factorization in terms of the so-called {\it holomorphic blocks}~\cite{Beem:2012mb} and/or {\it fibering operators}~\cite{Closset:2018ghr}. 

Perhaps more surprisingly, both concepts of localization in field space and factorization of the partition function have been shown to make sense in effective theories of supergravity arising in the low-energy limit of string or M-theory. The concept of supergravity localization was first put forward in~\cite{Dabholkar:2010uh,Dabholkar:2011ec,Dabholkar:2014wpa} and developed more formally in~\cite{Murthy:2015yfa,deWit:2018dix}. More recently, the classical on-shell action of minimal gauged 4d supergravity was shown to factorize as a sum of specific contributions~\cite{BenettiGenolini:2019jdz}, and a more involved gluing mechanism in terms of {\it gravitational building blocks} was uncovered in~\cite{Hosseini:2019iad} once abelian vector multiplets are included. It is our goal in this paper to extend these factorization results by proposing a general formula for the $\log$-corrections to the supergravity on-shell action, captured by a one-loop determinant calculation. Using holography beyond large $N$, our results translate into a prediction for the behavior of~$\log N$ corrections to the partition functions of a large class of 3d SCFTs.

In more detail, we systematically analyze the~$\log$-corrections in a large set of asymptotically locally Euclidean AdS$_4$ backgrounds\footnote{In fact the general formula we propose below is valid for {\it all} such backgrounds, modulo some cases that are so far excluded due to some technical assumptions. This will be discussed in due course.} in 4d~$\mathcal{N}=2$ gauged supergravity. We do so by using the aforementioned method of supergravity localization, combined with the factorization of the on-shell action in AdS$_4$/CFT$_3$. The authors of~\cite{BenettiGenolini:2019jdz} showed that the classical on-shell action (after a careful application of holographic renormalization) can always be factorized as a sum based on two types of contributions, depending on the fixed locus of the canonical isometry~$\xi$ generated by the Killing spinors preserved on the asymptotically locally AdS$_4$ (AlAdS$_4$) background. It turns out that~$\xi$ is nowhere vanishing on the asymptotic boundary and the fixed point set of~$\xi$ lies in the interior of the four-dimensional space with connected components that are either fixed points (NUTs), or fixed two-manifolds (Bolts). At the zeroes of~$\xi$ the Killing spinors always become chiral or anti-chiral, and we can denote the NUTs and Bolts with a subscript $\pm$ to reflect this. The fixed point set of~$\xi$ for a number of interesting explicit examples is given in Table~\ref{tab:example} so that the reader can build an intuition regarding the typical examples we consider here.\footnote{We stress that the four-dimensional fixed points are always inside the bulk and do not coincide with the fixed points of the canonical isometry on the boundary. In the cases where the bulk background is a Lorentzian black hole (BH), we find that the fixed point set of~$\xi$ is actually infinitely far away from the boundary in the throat near the horizon. In the other cases the fixed point set is at a finite but non-zero distance from the boundary.}

\begin{table}[ht]
	\begin{center}
		\setlength{\tabcolsep}{7pt}
		\renewcommand{\arraystretch}{1.3} 
		\begin{tabular}{ c || c | c | c |} \hline
		\multicolumn{1}{|c||}{AlAdS$_4$ space}	 & fixed point set of $\xi$ & 3d boundary & refs \\ \hline\hline
		\multicolumn{1}{|c||}{Squashed spheres} & 1 pt: centre of $\mathbb{H}_4$ & $S_b^3$ & \cite{Chamblin:1998pz,Martelli:2011fu,Martelli:2011fw,Kapustin:2009kz,Hama:2011ea}  \\ \hline
		\multicolumn{1}{|c||}{Static twisted BHs} & $\Sigma_\frak{g}$: centre of $\mathbb{H}_2$ & $S^1 \times \Sigma^\text{tw}_\frak{g}$ & \cite{Cacciatori:2009iz,Katmadas:2014faa,Halmagyi:2014qza,Benini:2015noa,Benini:2015eyy,Benini:2016hjo,Closset:2016arn,Benini:2016rke}  \\ \hline
		\multicolumn{1}{|c||}{Rotating twisted BHs} & 2 pts: NP/SP of $S^2$ $\times$ centre of $\mathbb{H}_2$ & $S^1 \times S^{2, \text{tw}}_\omega$ &  \cite{Hristov:2018spe,Benini:2015noa} \\ \hline
		\multicolumn{1}{|c||}{Kerr-Newman BHs} & 2 pts: NP/SP of $S^2$ $\times$ centre of $\mathbb{H}_2$ & $S^1 \times S^2_\omega$ & \cite{Caldarelli:1998hg,Hristov:2019mqp,Kim:2009wb,Imamura:2011su}  \\ \hline
		\multicolumn{1}{|c||}{1/4-BPS Bolts} &  $\Sigma_\frak{g}$: ${\cal O} (-{\frak p}) \rightarrow \Sigma_{\frak g}$ in $\mathbb{H}_4$ & $\cal{M}_{\frak{g}, \frak{p}}$ &  \cite{Closset:2017zgf,Toldo:2017qsh,Closset:2018ghr} \\ \hline
		\end{tabular}
	\end{center}
	\caption{The fixed point set of $\xi$ for a number of (non-singular) supersymmetric asymptotically locally Euclidean AdS$_4$ (i.e.\ $\mathbb{H}_4$) solutions, and the corresponding three-dimensional holographically dual backgrounds. The first entry generically denotes a family of solutions which consist of various ways one can squash the boundary three-sphere. The suggested references discuss in detail either the 4d bulk background, the 3d partition function, or both.}
	\label{tab:example}
\end{table}

From an off-shell supergravity perspective, we can use the same canonical isometry to set up a localization calculation. This approach involves the study of one-loop determinants for arbitrary matter multiplets coupled to gauged supergravity. We will argue in this work that, for holographic purposes, the logarithmic corrections to the classical on-shell action receive contributions that also factorize into a sum over the fixed locus of~$\xi$ for \emph{all} supersymmetric AlAdS$_4$ backgrounds. The most elegant way of evaluating the relevant one-loop determinants is by computing the index of differential operators acting on quadratic fluctuations around the localization locus using the Atiyah-Singer index theorem and its specialization, the Atiyah-Bott formula. See~\cite{Atiyah:1974obx,shanahan2006atiyah} for a comprehensive introduction of the mathematical background. Our final result is therefore a {\it quantum} one-loop analog of the {\it classical} factorization of the on-shell action of~\cite{BenettiGenolini:2019jdz}. We emphasize that our analysis here only uses the intermediate result of~\cite{BenettiGenolini:2019jdz} regarding the fixed point set of the canonical isometry~$\xi$ which also holds {\it off-shell}, and therefore we are able to add arbitrary matter supermultiplets in our analysis. This is in contrast to the results of~\cite{BenettiGenolini:2019jdz} regarding the {\it on-shell} action that only holds in minimal supergravity and needs to be supplemented by a set of gluing rules when extra matter is present, as in the example of matter-coupled black holes discussed in~\cite{Hosseini:2019iad}. 

We propose the following holographic formula for the one-loop contribution to the supergravity on-shell action\footnote{In the formula below and everywhere else in the text we work in the {\it grand-canonical ensemble} of fixed chemical potentials, even if the boundary conditions for certain backgrounds allow for other options. See below~\eqref{eq:logZsaddle} for more details on this point.} on a regular 4d supersymmetric background $M_4$,
\begin{equation}
\begin{split}
\label{eq:main}
	- \log Z^\text{sugra}_\text{1-loop} (M_4) =&\; \alpha\,\Bigl(\frac{n_{f}}{2} + \sum_i\,(1 - \frak{g}_i)\Bigr) \log L + \ldots \\
	=&\; \alpha\, \frac{\chi(M_4)}{2}\, \log L + \ldots \, ,
\end{split}
\end{equation}
where~$n_{f}$ is the number of fixed points of the canonical isometry~$\xi$ for the given background, and the sum runs over the fixed two-manifolds of genus~$\frak{g}_i$.\footnote{Here we exclude the case of a non-trivial line bundle of degree~$\frak{p}_i$ over the fixed two-manifolds corresponding to the last entry of Table~\ref{tab:example}. It leads to additional subtleties for the method we use, as discussed in more detail in section~\ref{subsec:Bolts} and Appendix~\ref{app:Albin}.} The ellipses denote terms that are subleading holographically, i.e. they do not scale logarithmically with the AdS$_4$ length scale~$L$. The part in brackets above is fully determined by the topology of the background~$M_4$ and equals its regularized Euler characteristic~$\chi(M_4)$ as shown in e.g.~\cite{Gibbons:1979xm}. In contrast, the dynamical information in~\eqref{eq:main} is entirely contained in the constant prefactor~$\alpha$ specific to the supergravity theory under consideration. As we will show, it can be interpreted as counting the various multiplets of the theory with a particular weight.
 
Let us stress that, to derive~\eqref{eq:main}, we do not need to assume anything about the (massless and massive) field content of the effective supergravity theory, nor do we need to specialize to a two-derivative theory. Our result is therefore valid in the presence of arbitrary\footnote{As mentioned above, we rely on the fact that the canonical isometry has a fixed point set of NUTs and Bolts. Strictly speaking, off-shell this has been shown explicitly only after a certain extra assumption on the form of the auxiliary fields~\cite{Bobev:2020egg,Genolini:2021urf} which covers all possible two-derivative theories, but only a subset of the four- and higher-derivative theories. It is however natural to expect that the fixed-point set of the canonical isometry remains the same after relaxing this technical assumption and therefore our results should hold for all higher-derivative theories.} perturbative corrections to the on-shell action of the background~$M_4$ (see~\cite{Bobev:2020egg,Bobev:2020zov,Bobev:2021oku,Genolini:2021urf}). We will also show that the form of our one-loop result is valid, at leading order holographically, for {\it any} supersymmetric field content, making our formula applicable to a completely arbitrary set of supermultiplets that might not even be known at present. Although at first sight this generality may seem mathematically implausible, we stress once more that we restrict our analysis to a very specific set of supersymmetric backgrounds~$M_4$, constrained both by supersymmetry and by the AlAdS$_4$ boundary conditions. This restriction ensures that we only need to apply the index theorem to the cases of zero- and two-dimensional fixed submanifolds, which in turn substantially simplifies  the mathematical problem and ultimately leads to \eqref{eq:main}.

Holographically, the length scale~$L$ of AdS$_4$ (taken to be large with respect to the higher dimensional length scale $l_P$ in order to stay within the supergravity approximation) is related to the rank~$N$ of the gauge group of the dual field theory as
\be
\label{eq:holorel}
	N \propto (L/l_P)^s\ ,
\ee
for some positive integer $s$ that is model-dependent. Our main supergravity formula~\eqref{eq:main} therefore translates into the following holographic prediction for the~$\log N$ corrections to the partition function of \emph{any} 3d~$\cN = 2$ SCFT with a supergravity dual,
\be
\label{eq:FTmain}
	\log Z^\text{SCFT}_{\partial M_4} (\frak{a}, N) = {\cal P} (\frak{a}, N) - \frac{\alpha}{s}\, \frac{\chi(M_4)}{2}\, \log N + \ldots \, ,
\ee
where~$\frak{a}$ denotes the collection of equivariant parameters pertaining to a given boundary background $\partial M_4$ together with possible discrete parameters (\textit{e.g.} Chern-Simons level, topological charges of the background),~$\cal{P}$ denotes the complete set of perturbative (polynomial in~$N$) corrections in a given partition function, and the ellipses denote possible non-perturbative (exponentially suppressed in~$N$) corrections. One concludes that the $\log N$ term does {\it not} depend on any of the equivariant parameters of the partition function. The above formula, based on arguments coming from supergravity localization, is expected to hold for \emph{any} supersymmetric partition function of a given holographic 3d SCFT. It is particularly easy to infer from~\eqref{eq:FTmain} that the logarithmic correction to the (refined) topologically twisted index is exactly equal to the one of the superconformal index, and exactly twice as much as the logarithmic correction for various squashed~$S_b^3$ partition functions. This follows from the fact that the latter backgrounds correspond to a single fixed point ($\chi = 1$) while the rotating black holes have two fixed points ($\chi = 2$) from our bulk four-dimensional perspective, see Table~\ref{tab:example} and \cite{Bobev:2020egg}. It is also clear from~\eqref{eq:FTmain} that the constant~$\alpha$, or more precisely the ratio~$\alpha / s$, should be interpreted as counting in a specific way the degrees of freedom of the holographic 3d SCFT. It would be very interesting to understand the above formula from a purely field theoretic point of view and find an independent definition of this coefficient.

At this point we should also stress one major caveat of our approach. We will show that the different supergravity multiplets will each give some specific contribution to the constant $\alpha$ in \eqref{eq:main}, but we cannot at present obtain an explicit derivation of the value of $\alpha$ for string theory compactifications of holographic interest. This is because we are so far unable to evaluate the contribution to $\alpha$ from {\it all} possible supermultiplets. This also means that we are unable to perform the necessary summation over the full Kaluza-Klein (KK) tower of multiplets present in the effective 4d description of a given string theory compactification. On the other hand, the strength of our approach is that once this single constant~$\alpha$ is known, we readily obtain the~$\log$-correction to all possible BPS backgrounds. The explicit value of $\alpha$ can in practice be fixed by a single known $\log$-correction: a holographic match with $\log N$ localization results in the dual field theory; or from the so-called {\it heat kernel} method applied in a careful way in supergravity. Such supergravity calculations were first presented for asymptotically flat BPS black holes in 4d~\cite{Banerjee:2010qc,Banerjee:2011jp,Sen:2011ba,Sen:2012cj},\footnote{Note that the heat kernel method is also well-suited for studying the $\log$-correction in the absence of supersymmetry as in~\cite{Sen:2012dw,Charles:2015eha,Castro:2018hsc,Karan:2021teq} and references therein. In contrast, our approach crucially relies on supersymmetry.} and an exemplary 11d supergravity calculation was conducted in~\cite{Bhattacharyya:2012ye} for empty AdS$_4$ uplifted on~$S^7$. This approach was more recently revisited in \cite{Liu:2017vbl} and further used in~\cite{Gang:2019uay,Benini:2019dyp}, and we give a more detailed account of it in Appendix \ref{app:11d}. From our point of view, these methods provide a complementary calculation of the~$\log$-correction that should agree with our general result, but are practically better suited for determining explicitly the constant $\alpha$ in a given compactification rather than deriving from first principles the factorization formula \eqref{eq:main}.

Based on both field theory and 11d calculations available in the literature, we can present some explicit values for the constant~$\alpha$ in several interesting holographic examples, see Table~\ref{tab:coeff}. We should note that, in the cases where we can infer the constant $\alpha$ from multiple independent calculations, we find complete agreement with~\eqref{eq:main}. A notable exception is the last entry (see \cite{Schwarz:2004yj,Guarino:2015jca}) where numerical studies of the topologically twisted index \cite{Liu:2018bac} predict $\alpha = 7/3$ while numerical studies of the $S^3$ partition function \cite{Liu:2019tuk} predict $\alpha = 8/3$. We expect further analysis to reconcile this apparent discrepancy.

\begin{table}[ht]
	\begin{center}
		\setlength{\tabcolsep}{7pt}
		\renewcommand{\arraystretch}{1.3}
		\begin{tabular}{| c | c || c | c | c |} \hline
	 3d SCFT & 4d SUGRA & $\alpha$ & $s$ & refs \\ \hline \hline
		 M2 branes on $Cone ({\cal M}_7)$ & 11d on ${\cal M}_7$ & $3$ & $6$ & \cite{Marino:2011eh,Fuji:2011km,Bhattacharyya:2012ye,Hatsuda:2016uqa,Liu:2017vll,Chester:2018aca,PandoZayas:2020iqr}  \\ \hline
	 $\SU(N)\, \, 6\mathrm{d}\, \, (2,0)$ on ${\cal M}_3$ & 11d on $S^4 \times {\cal M}_3$ & $3 (1-b_1^{{\cal M}_3})$ & $3$ & \cite{Gang:2018hjd,Gang:2019uay,Benini:2019dyp} \\ \hline
 $\SO(2 N)\, \, 6\mathrm{d}\, \, (2,0)$ on ${\cal M}_3$ & 11d on $\mathbb{RP}^4 \times {\cal M}_3$ & $0$ & $3$ & \cite{Bobev:2020zov} \\ \hline \hline
 ABJM at level $k \sim N$  & 10d IIA on $\mathbb{CP}^3$ & $1$ & $3$ & \cite{Marino:2011eh,Hanada:2012si}  \\ \hline
 $\U(N)^2$ quiver at level $k \sim 1$  & 10d mIIA on $\mathbb{CP}^3_\text{def}$ & $2$ & $6$ & \cite{Hong:2021bsb} \\ \hline
$\SU(N)$ at level $k \sim 1$ & 10d mIIA on $S^6_\text{def}$ & $7/3$ or $8/3$ & $6$ & \cite{Liu:2018bac,Liu:2019tuk} \\ \hline
		\end{tabular}
	\end{center}
	\caption{The constants $\alpha$ and $s$ in \eqref{eq:FTmain} for several classes of 3d SCFTs and the corresponding supergravity compactifications to an effective 4d model, together with the appropriate references. The first entry covers a range of M2 brane theories including ABJM~\cite{Aharony:2008ug} along with other $\cN=3$ and $\cN=4$ quivers~\cite{Imamura:2008nn,Jafferis:2008qz,Mezei:2013gqa}, and even a few examples where ${\cal M}_7$ is a Sasaki-Einstein manifold preserving only $\cN=2$ supersymmetry~\cite{Marino:2011eh,Hosseini:2016ume,PandoZayas:2020iqr}.}
	\label{tab:coeff}
\end{table}

The rest of this paper is organized as follows. In the next section we discuss the general set-up for calculating the one-loop determinant in the framework of supergravity localization, and state the main assumptions that go in the derivation of the result. In Section \ref{sec:vectors} we present the calculation of the index governing the one-loop determinant for abelian vector multiplets. We split the discussion into the two main cases of fixed 2-manifolds and fixed points in Sections \ref{subsec:Bolts} and \ref{subsec:NUTs}, respectively. In Section \ref{sec:multiplets} we present the general argument leading to the form of the one-loop correction for an arbitrary multiplet, and more briefly comment on certain supergravity multiplets such as hypermultiplets and the Weyl multiplet. We conclude with a list of remarks and related open problems in Section \ref{sec:conclusion}. Some of the more technical steps used in our analysis are relegated to the appendices. In Appendix \ref{app:Albin} we discuss more carefully the mathematical framework behind the regularization of the one-loop determinant on non-compact spaces with boundaries. In Appendix \ref{app:11d} we explain how these results relate to the eleven-dimensional approach used in the literature. Appendix \ref{app:Shanahan} contains the details of the computation of the equivariant index for vector multiplets on Bolt backgrounds using the Atiyah-Singer theorem.

%%%%%%%%%%%%%%%%%%%%%%%%%%%%%%%%%%%%%
\section{The one-loop set-up}
\label{sec:setup}
%%%%%%%%%%%%%%%%%%%%%%%%%%%%%%%%%%%%%

Before discussing the computation of one-loop determinants, we briefly review the steps one needs to perform in order to localize the supergravity path-integral. To a large extent, these steps coincide with the ones taken in the usual rigid localization algorithm, as reviewed in e.g. \cite{Pestun:2016zxk}. There are however some crucial differences that were highlighted in~\cite{deWit:2018dix}. In particular, in order to deal with the fact that one must integrate over the metric in the path-integral, the starting point is to prescribe a set of \emph{boundary conditions} for all the fields in the theory. This amounts to a choice of background, which we can conveniently take to be a supersymmetric \emph{solution} to the supergravity equations of motion. Once this solution is specified, we can build a suitable background field formalism in which we integrate over quantum fluctuations that vanish at the asymptotic boundary.

In the case of interest to us, the holographic dictionary further requires us to study asymptotically locally AdS$_4$ background solutions, for which there exists a well-defined procedure to extract finite observables \cite{Emparan:1999pm,Skenderis:2002wp}. Importantly, we will work in Euclidean signature throughout, which allows us to include all supersymmetric Lorentzian backgrounds (after a suitable Wick-rotation) as well as genuinely Euclidean solutions of holographic interest. As reviewed in \cite{BenettiGenolini:2019jdz}, based on the original analysis of \cite{Dunajski:2010uv}, any asymptotically locally AdS$_4$ supersymmetric background in Euclidean signature possesses a $\U(1)$ isometry defined by the Killing vector $\xi$ with the special property
\be
\label{eq:mainKStoKV}
	\xi^\mu = \bar{\varepsilon} \gamma^\mu \varepsilon\, ,
\ee
where $\varepsilon$ is a Killing spinor. One can always introduce local coordinates using this isometry and, as shown in \cite{BenettiGenolini:2019jdz} (building on the older ideas in \cite{Gibbons:1979xm}), one can completely determine the holographically renormalized on-shell action solely from the knowledge of the fixed points of the Killing vector $\xi$. It is also straightforward to derive the Killing vector $\xi$ from the off-shell gravitino variation, see \cite{Genolini:2021urf}, which is explicitly used in our analysis here. 

Having specified the background, we can interpret the corresponding Killing spinor~$\varepsilon$ as a ghost for the local supersymmetry transformations in the BRST formalism. At the same time, we also introduce ghost fields for all other local gauge transformations, e.g. due to the presence of vector or tensor multiplets. Deforming the resulting BRST algebra as in~\cite{deWit:2018dix} produces an equivariant supercharge~$Q$ whose algebra closes \emph{off-shell} according to
\be\label{eq:equivQ}
	Q^2 = {\cal L}_\xi \, ,
\ee
where~$\mathcal{L}_\xi$ is the Lie derivative along the Killing vector~$\xi$. Defined in this way, the charge~$Q$ leaves the background fields invariant while acting non-trivially on the quantum fluctuations. It can therefore be used to localize the supergravity path-integral.\footnote{Observe that in the case where the background has many Killing spinors, one can make a choice as to which~$Q$ to use to localize the path-integral.} Owing to the presence of~$\xi$ in~\eqref{eq:equivQ}, this path-integral localization will be related to the fixed-point formula~\cite{BenettiGenolini:2019jdz} for the on-shell action. Below, we will make this relation explicit also at the level of the one-loop determinants.
 
To localize the path-integral, we proceed as in the rigid case and deform the supergravity action by a~$Q$-exact term,~$S^\text{sugra} \rightarrow S^\text{sugra} + \lambda\,Q\mathcal{V}$. The fermionic functional~$\mathcal{V}$ is chosen so that~$Q^2\mathcal{V} = 0$, and this ensures that the deformed path-integral is in fact~$\lambda$-independent by the usual argument.\footnote{We assume here that the supergravity theory is free of local anomalies so that the field space measure is also invariant under~$Q$.} In the~$\lambda \rightarrow \infty$ limit, the supergravity path-integral thus localizes to the critical points of the deformation. A convenient choice for~$\mathcal{V}$ is to take
\be
\mathcal{V} = \int d^4 x \sum_{\Psi} \sqrt{g}\,\bar{\Psi} Q\Psi \, ,
\ee
where~$g$ is the determinant of the background metric and the sum runs over all quantum fluctuations of the fermion fields (which have vanishing background expectation values). The bosonic critical locus then consists of fields~$\phi$ that satisfy~$Q\Psi = 0$ and asymptote to the boundary values specified by our choice of background. Note that this locus splits into a product over loci in each multiplet of the theory owing to the fact that~$Q$ descends from the \emph{off-shell} local supersymmetry (and BRST) algebra.

With this notation and in the~$\lambda \rightarrow \infty$ limit, the path-integral reduces to 
\be\label{eq:locZ}
	Z^\text{sugra} (M_4) = \int [ {\rm d} \phi ]\, \exp\bigl(-S^\text{sugra} (\phi)\bigr)\,Z^\text{sugra}_{\text{1-loop}} (\phi)\, ,
\ee
where the one-loop contribution comes from integrating over the quadratic fluctuations of the fields around the localization locus, and one should carefully understand the measure of the remaining integral, as dictated by consistency of the quantum theory (see~\cite{LopesCardoso:2006ugz,Denef:2007vg,Murthy:2015yfa} for discussions on this point). On the left-hand side, we indicate explicitly that the localized path-integral depends on a choice of asymptotically locally AdS$_4$ background, which we denote by~$M_4$.

Our interest in the present paper does not lie in the complete evaluation of the localized path-integral~\eqref{eq:locZ} given a supersymmetric background~$M_4$, which at present seems out of reach. Instead, we are specifically interested in the structure of the one-loop factor~$Z^\text{sugra}_{\text{1-loop}}$ for arbitrary asymptotically locally AdS$_4$ supersymmetric backgrounds. In this context, we will not need the details of the localization locus specified by the fields~$\phi$. Indeed, we can focus on the evaluation of~\eqref{eq:locZ} in a saddle-point approximation around the classical background~$\phi = \mathring{\phi}(M_4)$, in which case the logarithm of the partition function reads
\be\label{eq:logZsaddle}
	\log Z^\text{sugra} (M_4) = - S^\text{sugra} \bigl(\mathring{\phi}(M_4)\bigr) + \log  Z^\text{sugra}_{\text{1-loop}} \bigl(\mathring{\phi}(M_4)\bigr) + \ldots
\ee
The first term on the right-hand side is the supersymmetric on-shell action, while the second term will scale as~$\log L$ with~$L$ the length scale associated to the asympotically locally AdS$_4$ background~$M_4$. Holographically, this length scale is related to the rank of the gauge group as in~\eqref{eq:holorel}, and we can therefore focus solely on the second term in~\eqref{eq:logZsaddle} to understand the~$\log N$ corrections to the planar limit in the dual field theory. Further corrections to the saddle-point approximation in~\eqref{eq:logZsaddle} are expected to go as~$N^{-n}$ for some (model-dependent)~$n\geq0$ and will not contribute to the log correction. Importantly, note that the above formula does {\it not} include the correction to the saddle-point evaluation of the classical action, which could potentially produce another~$\log N$ piece holographically as in the example of twisted black holes discussed in \cite{Hristov:2019xku}. Such corrections are not generic and do not correct the {\it grand-canonical} expression for the localized observable, but clearly need to be kept in mind. In holographic matches of observables in different ensembles one therefore needs to take extra care in the passage between the gravitational log-corrections~\eqref{eq:main} and the~$\log N$ field theory expression~\eqref{eq:FTmain}. 

Having reviewed the supergravity localization steps, we focus on the one-loop determinant of the quadratic fluctuations around the localization locus. This factor is most efficiently studied by introducing the so-called cohomological split with respect to the supercharge~$Q$. This means that we divide all bosonic and fermionic fields into sets~$\{ \mathbb{X}^{\cal X}_0, \mathbb{X}^{\cal X}_1 \}$ together with their~$Q$-images, where the index~${\cal X}$ runs over all supermultiplets in the theory,~$0$ denotes bosonic fields and~$1$ fermionic fields. The fermionic deformation~$\mathcal{V}$, at quadratic order around the localization locus, can be written as follows \cite{Pestun:2007rz}:
\be 
\label{eq:deformation}
\mathcal{V}|_{\text{quad.}} = \sum_{\cal X}\,\bigl(Q\mathbb{X}^{\cal X}_0 \;\; \mathbb{X}^{\cal X}_1\bigr) \; \begin{pmatrix} D_{00} & D_{01} \\ D_{10} & D_{11} \end{pmatrix}
\begin{pmatrix} \mathbb{X}^{\cal X}_0  \\ Q\mathbb{X}^{\cal X}_1 \end{pmatrix} \, .
\ee
The only quadratic modes that are not automatically paired by supersymmetry are the ones acted upon by the~$D_{10}$ operator. Therefore, for a given eigenvalue of $H := Q^2$, we are after the dimensions of the kernel and cokernel of~$D_{10}$. The difference of these dimensions is encoded in the equivariant index, 
\bea
\label{eq:indD10}
\begin{split}
\text{ind}_H(D_{10})(t) :=&\; \text{Tr}_\text{Ker$D_{10}$}  \, e^{\mathrm{i}Ht} - \text{Tr}_\text{Coker$D_{10}$}  \, e^{\mathrm{i}Ht} \\[1mm] 
=&\; \sum_{\mathfrak{n}}\,(m_{\mathfrak{n}}^{(0)} - m_{\mathfrak{n}}^{(1)})\,e^{\lambda_{\mathfrak{n}}t} \, ,
\end{split}
\eea
where~$m_{\mathfrak{n}}^{(0)}$ and~$m_{\mathfrak{n}}^{(1)}$ are the dimensions of the kernel and cokernel for a given eigenvalue~$\lambda_{\mathfrak{n}}$ of the~$\mathrm{i}H$ operator labeled by a set of parameters~$\mathfrak{n}$, and~$t$ is a formal expansion parameter. The equivariant index of differential operators is a subject of rich mathematical interest culminating in the Atiyah-Singer index theorem~\cite{atiyah1963index}. Broadly speaking, the theorem relates the number of solutions of the differential equations~$D_{10}u = D_{10}^*u = 0$ (where~${}^*$ denotes the formal adjoint) to the topology of the supersymmetric background and allows us to compute the index rather straightforwardly. Once the latter is known, the one-loop factor in~\eqref{eq:locZ} is obtained by making use of the following relation:
\begin{equation}
\label{eq:1-loop-formal}
Z^\text{sugra}_\text{1-loop} = \prod_{\mathfrak{n}}\,\lambda_{\mathfrak{n}}{}^{\frac12\, \frak{m}_{\mathfrak{n}} } \, , \qquad \frak{m}_{\mathfrak{n}} := m_\mathfrak{n}^{(1)} - m_\mathfrak{n}^{(0)} \, .
\end{equation}
Note that the infinite product over the~$\mathfrak{n}$ labels is a priori only a formal expression, and one may need to introduce a suitable regulator to obtain a finite, sensible answer. \\

After these general preliminaries, we proceed to the derivation of our main formula~\eqref{eq:main} for the case of vector multiplets on a number of supersymmetric backgrounds. We then give arguments for the general case of arbitrary supermultiplets. It is important to stress that we make three major assumptions in the derivation, together with a few smaller technical assumptions we discuss along the way:
\begin{itemize}
 \item we assume a {\it smooth filling} for all asymptotic boundaries we consider. We do not discuss backgrounds with singularities which possibly evade many of the arguments we present. Supergravity methods in general seem ill-suited for dealing with such singular spaces, which instead can often be resolved and understood in the framework of string theory.\footnote{In particular this means we also discard the supersymmetric black holes with spindle horizons recently discussed in \cite{Ferrero:2020twa} due to the appearance of conical singularities in the four-dimensional description. The existence of a gluing/fixed-point formula also for these horizons \cite{Hosseini:2021fge,Cassani:2021dwa} however strongly hints at a similar factorization of the one-loop determinant.}

\item  we assume a {\it choice of boundary conditions} for all supergravity fields on the asymptotically locally Euclidean AdS$_4$ backgrounds, such that we can use the Atiyah-Singer index theorem and the Atiyah-Bott fixed point formula as stated for compact spaces. Such an assumption is intrinsically at the heart of holography beyond large $N$ and the index theorems we use are well-studied for non-compact spaces with boundaries. We therefore comment in more detail on the issues of regularization on non-compact spaces below and in Appendix \ref{app:Albin}, based on the analysis of \cite{albin2007renormalized}.

\item  we consider only {\it abelian} gauge groups in supergravity. Although this is not a major loss of generality as far as holography is concerned (it means that the underlying R-symmetry group of the dual field theory can only be~$\U(1)$ such that we describe the holographic 3d SCFT in an~$\cN = 2$ language), it simplifies the gauge-fixing using BRST as well as the calculation of the equivariant index of differential operators.
\end{itemize}

From an AdS/CFT point of view these three main assumptions fit in the principles of standard holography and are certainly not surprising, but it would be interesting to explore more rigorously their fundamental importance. We leave this question for future studies.

%%%%%%%%%%%%%%%%%%%%%%%%%%%%%%%%%%%%%
\section{Vector multiplets}
\label{sec:vectors}
%%%%%%%%%%%%%%%%%%%%%%%%%%%%%%%%%%%%%

Here we follow closely the logic and detailed calculations presented in~\cite{Hristov:2019xku} for the example of supersymmetric black holes in AdS$_4$, extending it in several ways for the purposes of our general argument. We use the superconformal gravity formulation reviewed e.g. in~\cite{Lauria:2020rhc}. The main advantage is that it guarantees off-shell closure of the local symmetry algebra on the various multiplets, and ensures that the following analysis is independent of the precise form of the supergravity action, which can for instance include higher-derivative terms. The minimal gauged supergravity theory in this formalism consists of the gravity (Weyl) multiplet, one compensating vector multiplet and one compensating hypermultiplet. We can then add an arbitrary number of vector and hypermultiplets to obtain matter-coupled $\cN = 2$ gauged supergravities. In addition, there are more exotic multiplets available in the formalism like tensor multiplets and more general linear and chiral multiplets~\cite{deWit:1980lyi}, with various combinations of these also producing massive multiplets.

Let us now consider a number of abelian vector multiplets. The scalars and fermions are in the adjoint representation so we have no charged fields. Since the superconformal algebra closes on each multiplet separately, we can focus on the contribution of a single vector multiplet to the one-loop determinant without loss of generality. A vector multiplet in Euclidean superconformal gravity~\cite{deWit:2017cle} consists of two independent real scalar fields~$X_{\pm}$, a~$\U(1)$ gauge field~$W_\mu$, a triplet of real scalars~$Y_{ij}$, as well as a doublet of chiral and anti-chiral gaugini~$\Omega^i_\pm$. Here and below,~$i=1,2$ is an $\SU(2)_R$ index. As outlined above, we must gauge-fix the theory by introducing appropriate ghost fields and arrange all fields according to the cohomological split. To do so, it is convenient to change the basis for the physical fermion fields to the so-called twisted gaugini~$\lambda, \lambda_\mu, \lambda^{ij}$ as in~\cite{Hristov:2019xku}. Because the norm of the Killing spinor generating the canonical isometry~$\xi$ is non-vanishing, this change of basis is invertible. Next we can easily gauge-fix the~$\U(1)$ gauge symmetry by introducing a set of ghost and anti-ghost fields~$c$ and~$b$ together with a Lagrange multiplier field~$B$ enforcing the gauge-fixing condition in the path-integral (there will be one of these for each vector multiplet). Together with the local supersymmetry ghosts, this allows us to deform the standard BRST operator into an equivariant supercharge~$Q$~\cite{deWit:2018dix}, which closes off-shell according to~\eqref{eq:equivQ}. The cohomological split for a vector multiplet is then given by\footnote{See~\cite{Hristov:2019xku} for a discussion of the reality conditions obeyed by the fields.}
\begin{equation}
\label{eq:vec-split}
\mathbb{X}^\text{vec}_0 := \{ X_+ + X_-\,, W_\mu\} \, , \quad \mathbb{X}^\text{vec}_1 := \{\lambda^{ij}\,,c\,,b\} \, ,
\end{equation}
while the remaining bosonic fields~$\{X_+ - X_-, Y_{ij}, B\}$ and fermionic fields~$\{\lambda, \lambda_\mu\}$ sit in the~$Q$-images of the above sets. Combined with the expression for the fermionic deformation functional~$\cal{V}$, this split allows us to explicitly identify the~$D_{10}$ operator relevant for vector multiplets according to \eqref{eq:deformation}. Schematically, its matrix elements take the form
\begin{equation}
\label{eq:D10-schem}
(D^\mathrm{vec}_{10})_{\alpha\beta} = K^{\mu\nu}_{\alpha\beta}(\xi)\,\partial_\mu\partial_\nu + K^\mu_{\alpha\beta}(\xi)\,\partial_\mu + K_{\alpha\beta}(\xi) \, , 
\end{equation}
where~$\alpha,\beta = 1,\ldots,5$ run over the fields of~$\mathbb{X}^\text{vec}_{0,1}$. The derivative coefficients depend on the choice of supersymmetric background~$M_4$, and therefore on the choice of the canonical isometry~$\xi$. We refer the reader to~\cite{Hristov:2019xku} for explicit expressions in the case of static twisted black holes mentioned in Table \ref{tab:example}. \\

We now want to compute the equivariant index under $H = Q^2$ of the operator $D^\mathrm{vec}_{10}$. An important property of this index is that it is uniquely determined\footnote{This fact is true for so-called elliptic operators. For more general transversally elliptic operators, the index is determined by its symbol up to a distribution \cite{Atiyah:1974obx}.} by the \emph{symbol} of the differential operator. In practice, the symbol~$\sigma(D^\mathrm{vec}_{10})$ is obtained by replacing the derivatives~$\partial_\mu$ by~$p_\mu$ in~\eqref{eq:D10-schem}, where~$p_\mu$ are the coordinates on the fibers of the cotangent bundle~$T^*\!M_4$ (the momenta). This yields the following matrix representation of the symbol, 
\begin{equation}
\sigma(D^\mathrm{vec}_{10})_{\alpha\beta} = K_{\alpha\beta}^{\mu\nu}(\xi)\,p_\mu p_\nu + K_{\alpha\beta}^\mu(\xi)\,p_\mu \, .
\end{equation}
Using some linear algebra, it is then possible to show that~$\sigma(D^\mathrm{vec}_{10})$ is equivalent to the symbol of standard differential operators depending on whether the background~$M_4$ is a NUT or a Bolt. For the former, the isometry~$\xi$ has isolated fixed points and the symbol reduces to the symbol of the (anti)-self-dual complex
\begin{equation}
\label{eq:ASD}
D_\pm \; : \; 0 \longrightarrow \Omega^0 \stackrel{d}{\longrightarrow} \Omega^1 \stackrel{d^\pm}{\longrightarrow} \Omega^{2\pm} \longrightarrow 0 \, ,
\end{equation}
in the~$\mathbb{R}^4$ neighborhood of each fixed point~\cite{Pestun:2007rz}. For the latter,~$\xi$ leaves invariant a two-dimensional submanifold isomorphic to a Riemann surface and the symbol of $D^\mathrm{vec}_{10}$ can be identified with the symbol of $D_+$ in the neighborhood of the fixed locus~\cite{Hristov:2019xku}. The upshot of this short discussion is that we can access the one-loop determinants of abelian vector multiplets coupled to gauged supergravity by evaluating the equivariant index of the standard complex~\eqref{eq:ASD} in the neighborhood of the fixed locus, both for NUT and Bolt backgrounds.

%%%%%%%%%%%%%%%%%%%%%%%%%%%%%%%%%%%%%
\subsection{Fixed 2-manifolds (``Bolts'')}
\label{subsec:Bolts}
%%%%%%%%%%%%%%%%%%%%%%%%%%%%%%%%%%%%%

We first consider the case of the canonical isometry~$\xi$ having a fixed two-dimensional submanifold, necessarily isomorphic to a Riemann surface~$\Sigma_\frak{g}$ of genus~$\frak{g}$.  The case of fixed two-manifolds instead of fixed points prominently features in the rigid localization of the topologically twisted index~\cite{Benini:2015noa,Benini:2016hjo,Closset:2016arn}, but below we will employ an index theorem instead of the direct mode analysis used in these references. The neighborhood of the fixed two-manifold is most generally given by a complex line bundle~${\cal O} (-{\frak p}) \rightarrow \Sigma_{\frak g}$ with the fibration having Chern degree~$\frak{p}$. Locally we write the canonical Killing vector as
\be
\label{eq:KVBolt}
	\xi = \frac{1}{L_b}\, \partial_\tau\ , \qquad \xi^\mu =: (\varepsilon ,0,0,0) \, ,
\ee
where~$\tau$ is the polar angle covering the complex line bundle (Euclidean time). We have inserted the length scale~$\varepsilon^{-1} = L_b$ on dimensional grounds, and note that the particular value of~$L_b$ in general depends (i) on the particular background~$M_4$ and (ii) on the localization locus parametrized by $\phi$ in~\eqref{eq:locZ}. In the case of static twisted black holes \cite{Cacciatori:2009iz} where the fibre is trivial and the fixed two-manifold is located at the centre of the~$\mathbb{H}_2$ factor in the near-horizon geometry,~$L_b$ measures the \emph{off-shell} length scale of~$\mathbb{H}_2$~\cite{Hristov:2019xku}; in the case of the 1/4-BPS Bolt solutions with a general fibre~$\frak{p}$~\cite{Toldo:2017qsh} or the black saddles with trivial fibre \cite{Bobev:2020pjk}, the fixed two-manifold is embedded in~$\mathbb{H}_4$ and~$L_b$ measures the length scale of the asymptotic~$\mathbb{H}_4$. No matter what the particular details are, holographically we are interested in the scale carrying the rank of the gauge group~$N$. We will discuss its relation to~$L_b$ at the end of this subsection.

%%%%%%%%%%%%%%%%%%%%%%%%%%%%%%%%%%%%%
\subsubsection*{Non-compactness}
\label{subsubsec:non-compact}
%%%%%%%%%%%%%%%%%%%%%%%%%%%%%%%%%%%%%

We now want to use the Atiyah-Singer theorem to obtain the equivariant index of the self-dual complex~\eqref{eq:ASD}. The fact that we consider non-compact asymptotically locally Euclidean AdS$_4$ backgrounds introduces potential subtleties. To discuss such subtleties in a simple setting, recall that on an even-dimensional compact manifold~$M_4$ with a boundary, Chern's Gauss-Bonnet index formula states 
\begin{equation}
\int_{M_4} e+ \int_{\partial M_4} \Theta = \chi(M_4) \, .
\end{equation}
Above,~$e$ is the Euler class of~$M_4$ and~$\Theta$ is a polynomial in the curvature two-form~$\mathcal{R}^{ab}$ and the second fundamental form~$\theta^{ab}$. In four dimensions, they are given by the familiar expressions~\cite{Eguchi:1980jx}
\begin{equation}
\label{eq:4deT}
e = \frac{1}{32\pi^2}\,\varepsilon_{abcd}\,\mathcal{R}^{ab} \wedge \mathcal{R}^{cd} \, , \quad \Theta = -\frac{1}{32\pi^2}\,\varepsilon_{abcd}\Bigl(2\,\theta^{ab} \wedge \mathcal{R}^{cd} - \frac43\,\theta^{ab} \wedge \theta^c{}_e \wedge \theta^{ed}\Bigr) \, .
\end{equation}
Chern's formula can be generalized to the case of non-compact manifolds by introducing suitable regulators for the integrals, in order to deal with the potential divergences. We will not review this in detail here and simply refer the reader to~\cite{albin2007renormalized} for a thorough exposition. Of particular interest to us is the application of such a regularized Gauss-Bonnet formula to the case where the metric on the background~$M_4$ is \emph{conformally compact}. A conformally compact metric is one that, in a neighborhood of the boundary located in some coordinate system at~$x = 0$, can be put in the form
\begin{equation}
\label{eq:CCmetric}
ds^2 \approx \frac{dx^2}{\alpha(x)^2 x^2} + \frac{h_{ij}(x) dy^i dy^j}{x^2} \quad\; \text{as} \quad x \rightarrow 0 \, ,
\end{equation}
with~$\alpha(0) \neq 0$. If in addition the boundary metric~$h_{ij}(x)$ has an expansion around~$x=0$ involving only \emph{even} powers of~$x$ below~$x^4$, then it can be shown that 
\begin{equation}
{}^R\int_{\partial M_4} \Theta = 0 \, ,
\end{equation}
where ${}^R$ denotes an explicit regularization of the integral~\cite{albin2007renormalized}. Thus, for such class of metrics, the Gauss-Bonnet index formula receives no contribution from the boundary and takes the form of a simple regularization of the standard Gauss-Bonnet theorem applicable to closed manifolds,
\begin{equation}
{}^R\int_{M_4} e = \chi(M_4) \, .
\end{equation}
Motivated by this simple example, we will restrict our application of the Atiyah-Singer index theorem to supersymmetric backgrounds whose metric is of the above class, namely even conformally compact (ECC). In the case of the de Rham complex responsible for the Gauss-Bonnet theorem, this restriction addresses the potential issue of boundary contributions to the index, as we just saw. The issue of non-compactness can be dealt with by regularizing the integrals as in the usual holographic renormalization, namely by introducing a cut-off $x = \delta$, expanding around small $\delta$ and keeping only the constant term~\cite{albin2007renormalized}. It seems reasonable to expect that the restriction to ECC metrics will also avoid subtle boundary issues for other complexes. We will thus work under the assumption that the equivariant index of standard complexes is free of boundary contributions and can be regularized straightforwardly when the metric on~$M_4$ is ECC. Clearly, the Atiyah-Singer theorem may very well apply to backgrounds outside of this class, but we leave this as an interesting (mathematical) question for the future.

%%%%%%%%%%%%%%%%%%%%%%%%%%%%%%%%%%%%%%%%
\subsubsection*{Atiyah-Singer index theorem}
\label{subsubsec:AS}
%%%%%%%%%%%%%%%%%%%%%%%%%%%%%%%%%%%%%%%%

For Bolt backgrounds with a general fiber, we show in Appendix~\ref{app:Albin} that the metric is conformally compact but not necessarily even. The failure to be even is controlled by the Chern degree~$\mathfrak{p}$ of the fibration. According to the above discussion, we will therefore restrict ourselves to the case~$\mathfrak{p} = 0$ in what follows.\footnote{We make some comments about the $\mathfrak{p} \neq 0$ case in Appendix~\ref{app:Shanahan}.} For such Bolts, the neighborhood of the fixed locus of~$\xi$ is simply~$N \cong \mathbb{R}^2 \times \Sigma_{\mathfrak{g}}$. Under the~$\U(1)_\varepsilon$ group action generated by~$\xi$ in~\eqref{eq:KVBolt}, the (complexified) self-dual complex~\eqref{eq:ASD} is isomorphic to the Dolbeault complex twisted by the holomorphic~$\U(1)$-bundle~$V := \mathcal{O} \oplus K_N$ where~$\mathcal{O}$ is the trivial bundle and~$K_N$ is the canonical bundle over~$N$ \cite{friedman2013smooth}. This twisted Dolbeault complex is given by
\begin{equation}
\label{eq:twist-del}
\bar{\partial}_{V} \; : \; 0 \longrightarrow \Omega^{0,0}(V) \stackrel{\bar{\partial}_V}{\longrightarrow} \Omega^{0,1}(V) \stackrel{\bar{\partial}_V}{\longrightarrow}\Omega^{0,2}(V) \longrightarrow 0 \, .
\end{equation}
The Atiyah-Singer theorem then gives the equivariant index of the operator~$\bar{\partial}_{V}$ as an integral over the submanifold~$M_q \subset N$ left fixed by an element~$q \in \U(1)_\varepsilon$ \cite{shanahan2006atiyah}, 
\begin{equation}
\label{eq:equiv-dolb}
\mathrm{ind}_q(\bar{\partial}_{V}) = \int_{M_q}\,\frac{\mathrm{ch}_q(V|M_q)\,\mathrm{Td}(TM^{\,+}_q)}{\mathrm{ch}_q\bigl(1 - NM^{\,-}_q\bigr)}\Big\vert_{\mathrm{top}} \, ,
\end{equation}
with~$TM^{\,+}_q$ the holomorphic tangent bundle of~$M_q$,~$NM^{\,-}_q$ the anti-holomorphic normal bundle of~$M_q$ in~$M$, and~$V|M_q$ the restriction of the bundle~$V$ to~$M_q$. The characteristic classes of the relevant bundles appearing in the integrand are the Todd class (Td) and the $\U(1)$-equivariant Chern character (ch$_q$). Lastly, the subscript ``top'' indicates that we integrate the top-form component of the resulting class over $M_q$. 

In the case at hand,~$M_q \cong \Sigma_\mathfrak{g}$ and we can also distinguish between the orientation of the normal bundle~$N\Sigma_{\frak g}$ which is related to the chirality of the Killing spinors defining~$\xi$. This leads to two cases which~\cite{BenettiGenolini:2019jdz} dubbed Bolt$_\pm$. By the equivalence of symbols discussed above, the equivariant index~\eqref{eq:equiv-dolb} directly gives the equivariant index of the~$D_{10}^\text{vec}$ operator we are after. We relegate the evaluation of the relevant characteristic classes in the integrand to Appendix~\ref{app:Shanahan}, and simply quote the final result:
\begin{equation}
\label{eq:equiv-SD-C}
\mathrm{ind}_q(D_{10}^\mathrm{vec})(t) = -\Bigl(\frac{1}{1 - q} + \frac{1}{1 - q^{-1}} - 2\,\Bigr)\,\int_{\Sigma_\mathfrak{g}}\,\frac12\,c_1(\Sigma_\mathfrak{g}) \, .
\end{equation}
Above, $q := \exp(\mathrm{i}\,\varepsilon\,t) \in \U(1)_\varepsilon$ and the factor of $2$ in the first bracket was added by hand to remove the zero-modes, as explained in details in \cite{Hristov:2019xku}. We now expand the first geometric series in \eqref{eq:equiv-SD-C} in powers of~$q$ and the second series in powers of~$q^{-1}$. The equivariant index then reads
\begin{equation}
\label{eq:ind-vec-AS}
\mathrm{ind}_{\varepsilon}(D^\mathrm{vec}_{10})(t) = -\Bigl(\sum_{n = 1}^{\infty}\,e^{\mathrm{i}n t \varepsilon} + \sum_{n = 1}^{\infty}\,e^{-\mathrm{i}n t \varepsilon}\Bigr)\int_{\Sigma_\mathfrak{g}}\frac12\,c_1(\Sigma_\mathfrak{g}) = - \sum_{n\in\mathbb{Z}^*} \mathfrak{m}_n\,e^{\mathrm{i} n t \varepsilon} \, ,
\end{equation}
where the multiplicity is given by
\begin{equation}
\label{eq:multi-index}
\frak{m}_n = \int_{\Sigma_\mathfrak{g}}\,\frac12\,c_1(\Sigma_{\mathfrak{g}}) = 1 - \frak{g} \quad\;\; \forall\;n \in \mathbb{Z}^* \, .
\end{equation}
According to~\eqref{eq:KVBolt} we set~$\varepsilon = L_b^{-1}$ and use~\eqref{eq:1-loop-formal} to write down the one-loop determinant:
\begin{equation}
\label{eq:ZBolt-index}
Z^{\mathrm{vec}}_{\textnormal{1-loop}}(\mathrm{Bolt}_\pm) = \prod_{n \in \mathbb{Z}^*} \Bigl(\frac{\mathrm{i}n}{L_b}\Bigr)^{\mathfrak{m}_n/2} \, .
\end{equation}
Note that the factor of 2 in~\eqref{eq:equiv-SD-C} precisely removed the~$n=0$ zero-mode. The above result is in agreement with the direct mode computation in \cite{Hristov:2019xku} using the explicit form of the~$D_{10}^\text{vec}$ differential operator and imposing appropriate boundary conditions.

%%%%%%%%%%%%%%%%%%%%%%%%%%%%%%%%%%%%%%%%
\subsubsection*{Regularization and length scale}
\label{subsubsec:regularize}
%%%%%%%%%%%%%%%%%%%%%%%%%%%%%%%%%%%%%%%%

We must now regularize the infinite product~\eqref{eq:ZBolt-index}, and for this we use zeta-function regularization. As mentioned at the beginning of this subsection, we focus our attention on the scale carrying the rank~$N$ of the dual gauge group. This allows us to drop a number of irrelevant constants and concentrate on the~$L_b$-dependence alone. With this in mind, we write
\begin{equation}
\label{eq:Boltloop}
\log Z^{\mathrm{vec}}_{\textnormal{1-loop}}(\mathrm{Bolt}_\pm) = -\frac12\,\sum_{n\in\mathbb{Z}^*}\bigl(1 - \mathfrak{g}\bigr)\log L_b + \textnormal{pure number} = \frac12\,\bigl(1 - \frak{g}\bigr)\log L_b \, ,
\end{equation}
where we have used the analytic continuation~$\sum_{n\geq 1} n^0 = \zeta(0) = -1/2$. Note that we did not need the explicit form of the localization locus~$Q\mathcal{V}(\phi) = 0$ to arrive at~\eqref{eq:Boltloop}. The genus of the Riemann surface is a topological quantity and does not depend on the details of the particular supersymmetric background~$M_4$, or the associated localization locus. The one subtle point is that the length scale~$L_b$ is in principle a continuous parameter and can depend on such details. In the example of the static twisted black holes discussed in~\cite{Hristov:2019xku}, it was shown that~$L_b$ does vary along the localizing manifold. It is only in a saddle-point evaluation of the localized path-integral that~$L_b$ corresponds to the {\it on-shell} length of the~$\mathbb{H}_2$ factor in the near-horizon region, which in turn is related to the asymptotic length scale of~$\mathbb{H}_4$ via the black hole magnetic charges. Although it might seem that we cannot go further without analyzing the details of each specific case, this is not true for the purposes of analyzing specifically the logarithmic corrections to the partition functions of dual SCFTs. The idea is that inside the path-integral,~the $L_b$-dependence of the one-loop determinants gives a non-trivial integrand. However, we are only interested in the part of the one-loop contributions carrying holographically the rank~$N$ of the gauge group. In other words, we have not committed to fully evaluating any of the factors entering the ellipses in~\eqref{eq:main} or~\eqref{eq:FTmain}. These terms will come both from constant factors in the one-loop determinant, such as the pure number part in~\eqref{eq:Boltloop}, and from the finite dimensional localized integral over~$L_b(\phi)$. The latter can in principle generate many subleading terms in~$N$ that enter the perturbative polynomial $\cal{P}$ in~\eqref{eq:FTmain}. Since the logarithm separates such corrections, we can safely ignore these subtleties and replace the off-shell scale~$L_b$ in~\eqref{eq:Boltloop} with the on-shell length scale~$L$ that is related to the rank~$N$ via the holographic dictionary~\eqref{eq:holorel}. \\

We can now allow for an arbitrary number ($n_V + 1$) of abelian vector multiplets in a Bolt$_\pm$ background, corresponding to~$n_V$ physical vectors together with the conformal compensator introduced in the off-shell conformal supergravity formalism. Each such multiplet brings its own factor of~\eqref{eq:Boltloop} to the total one-loop factor. For completeness, we also allow for multiple fixed two-manifolds under the canonical isometry~$\xi$, even though all explicit examples we know of only have a single fixed~$\Sigma_\mathfrak{g}$. This way, we arrive at the final expression
\begin{equation}
\label{eq:Boltloopfinal}
\log Z^{\mathrm{vec}}_{\textnormal{1-loop}}(\mathrm{Bolt}_\pm) = \frac{n_V + 1}2\,\sum_i (1 - \frak{g_i})\log L + \ldots
\end{equation}
In the language of~\eqref{eq:main}, this shows that each vector multiplet in the gauged supergravity theory (physical or compensating) contributes~$-1/2$ to the constant~$\alpha$, 
\begin{equation}
\label{eq:alphavecBolt}
	\alpha^\mathrm{vec}_\text{Bolt} = -\frac12\ ,
\end{equation}
which we claimed could be thought of as a measure of the degrees of freedom in the bulk. We will expand on this point after analyzing the NUT case in the next subsection.

%%%%%%%%%%%%%%%%%%%%%%%%%%%%%%%%%%%%%%%%
\subsection{Fixed points (``NUTs'')}
\label{subsec:NUTs}
%%%%%%%%%%%%%%%%%%%%%%%%%%%%%%%%%%%%%%%%

%%%%%%%%%%%%%%%%%%%%%%%%%%%%%%%%%%%%%%%%
\subsubsection*{Atiyah-Bott fixed point formula}
\label{subsubsec:AB}
%%%%%%%%%%%%%%%%%%%%%%%%%%%%%%%%%%%%%%%%

Let us now look at the situation where the fixed locus of~$\xi$ is a set of isolated points. In this case, the Atiyah-Singer theorem simplifies further and we can directly use the Atiyah-Bott fixed point formula~\cite{Atiyah:1974obx}. We will begin by discussing the case of a single fixed point, as in the example of~$\mathbb{H}_4$ with a squashed 3-sphere~$S_b^3$ boundary, and then discuss the effect of having additional fixed points. For such a NUT, the neighborhood of the fixed point is just flat space,~$N \cong \mathbb{C}^2$. Locally the canonical Killing vector is given by
\be
\label{eq:KVNut}
	\xi = \varepsilon_1 \,\partial_{\tau_1} + \varepsilon_2 \,\partial_{\tau_2} = \frac1{L_b}\bigl(\partial_{\tau_1} + \omega\,\partial_{\tau_2}\bigr) \, , 
\ee
where~$\tau_{1,2}$ are the standard polar angles covering the two~$\mathbb{C}$-planes. In the second equality, we have again explicitly inserted an overall length scale~$L_b$ and used the dimensionless equivariant parameter $\omega$. The two~$\mathbb{C}$-planes are each associated with an~$\SU(2)$ Lorentz symmetry, which together make up the full~$\SO(4)$ symmetry group.

Every NUT background compiled in Table~\ref{tab:example} has a metric which is even conformally compact. According to our discussion in the previous subsection, we will therefore proceed and use the Atiyah-Bott fixed point formula to evaluate the equivariant index of the~$D_{10}^\text{vec}$ operator~\eqref{eq:D10-schem} under the~$G = \U(1)_{\varepsilon_1} \times \U(1)_{\varepsilon_2}$ action generated by~$\xi$. This formula reads
\begin{equation}
\label{eq:AB}
\mathrm{ind}_{q_1,q_2}(D_{10}^\mathrm{vec})(t) = \sum_{x\,|\,\widetilde{x}=x}\,\frac{\mathrm{Tr}_{\mathbb{X}^\text{vec}_0,\mathbb{X}^\text{vec}_1} (-1)^F\,e^{tH_{\varepsilon_1,\varepsilon_2}}}{\mathrm{det}\bigl(1-\partial\widetilde{x}/\partial x\bigr)} \, , \quad \text{where} \quad \widetilde{x} = e^{tH_{\varepsilon_1,\varepsilon_2}}\,x \, ,
\end{equation}
where~$q_{1,2} := \exp(\mathrm{i}\,\varepsilon_{1,2}\,t) \in G$, the sum is over the fixed points under the $G$-action~\eqref{eq:KVNut}, and the trace is taken over the field subspace consisting of the~$\mathbb{X}_0^\text{vec}$ and~$\mathbb{X}_1^\text{vec}$ sets that are not paired by supersymmetry (\textit{cf}.~\eqref{eq:vec-split}). The eigenvalues of~$H_{\varepsilon_1,\varepsilon_2}$ on this subspace are labeled by pairs of integers~$\mathfrak{n} = \{n_1,n_2\} \in \mathbb{Z}^2$,
\begin{equation}
\label{eq:refined-H}
H_{\varepsilon_1,\varepsilon_2}\mathbb{X}^\text{vec}_{0,1} = \bigl(\mathrm{i}\varepsilon_1 n_1 + \mathrm{i}\varepsilon_2 n_2\bigr)\mathbb{X}^\text{vec}_{0,1}  \, .
\end{equation}
The determinant factor in~\eqref{eq:AB} is given by the product
\begin{equation}
\label{eq:AB-denom}
\mathrm{det}\bigl(1-\partial\widetilde{x}/\partial x\bigr) = (1-q_1)(1-q_1^{-1})(1-q_2)(1-q_2^{-1}) \, .
\end{equation}
The numerator can be computed by analyzing the representation of the fields in~$\mathbb{X}^\text{vec}_{0,1}$ under the~$\SO(4)$ group. For the sets~\eqref{eq:vec-split}, the bosonic and fermionic traces are \cite{Hristov:2019xku}:
\begin{equation}
\begin{split}
\mathrm{Tr}_{\mathbb{X}^{\text{vec}}_0} (-1)^F\,e^{tH_{\varepsilon_1,\varepsilon_2}} =&\; 1 + q_1^{-1} + q_1 + q_2^{-1} + q_2 \, , \\
\mathrm{Tr}_{\mathbb{X}^{\text{vec}}_1} (-1)^F\,e^{tH_{\varepsilon_1,\varepsilon_2}} =&\; - 2 - q_1^{-1} q_2^{-1} - 1 - q_1 q_2 \, .
\end{split}
\end{equation}
Using this in~\eqref{eq:AB}, we arrive at the equivariant index
\be
\label{eq:NUT-q}
	\mathrm{ind}_{q_1,q_2}(D_{10}^\mathrm{vec})(t) = 1 -\frac{1 + q_1 q_2}{(1-q_1)(1-q_2)} \, ,
\ee
where we have added an extra constant to take care of the zero-mode, as in the Bolt case.

Depending on the orientation of the fixed point, i.e. whether we have a NUT$_+$ or a NUT$_-$ in the notation of~\cite{BenettiGenolini:2019jdz}, the Killing spinor at the fixed point has a fixed positive or negative chirality and one can accordingly think of~$\omega$ in~\eqref{eq:KVNut} as positive or negative. As a result, one should set up a consistent expansion for the geometric series in~\eqref{eq:NUT-q} in either positive or negative powers of~$q_2$ (without loss of generality we can always take positive powers in $q_1$). This is explained carefully in~\cite{Hosomichi:2016flq}. The resulting expansions are as follows:
\begin{equation}
\begin{split}
	\text{NUT}_+\;: \quad \mathrm{ind}_{q_1,q_2}(D_{10}^\mathrm{vec})(t) =&\; 1  - \sum_{n_1, n_2 \geq 0} (1+q_1 q_2)\, q_1^{n_1} q_2^{n_2} \\ 
	=&\;  - 2 \sum_{n_1, n_2 \geq 1} q_1^{n_1} q_2^{n_2} - \sum_{n_1 \geq 1} q_1^{n_1} -  \sum_{n_2 \geq 1} q_2^{n_2} \, ,
\end{split}
\end{equation}
and
\begin{equation}
\begin{split} 
	\text{NUT}_-\;: \quad \mathrm{ind}_{q_1,q_2}(D_{10}^\mathrm{vec})(t) =&\; 1 - \sum_{n_1, n_2 \geq 0} (1+q_1 q_2^{-1})\,  q_1^{n_1} q_2^{-n_2} \\ 
	=&\; -2 \sum_{n_1, n_2 \geq 1} q_1^{n_1} q_2^{-n_2} - \sum_{n_1 \geq 1} q_1^{n_1} - \sum_{n_2 \geq 1} q_2^{-n_2} \, .
\end{split}
\end{equation}
From this, we read off the~$D_{10}^\text{vec}$ eigenvalue multiplicities for the NUT$_+$,
\begin{equation}
\text{NUT}_+\;: \quad \mathfrak{m}_\mathfrak{n} = \begin{cases} 2 \quad &\text{for} \;\; (n_1, n_2) > (0,0) \\ 1 \quad &\text{for} \;\; (n_1>0,0) \;\; \text{or} \;\; (0,n_2>0) \\ 0 \quad &\text{otherwise} \end{cases} \, .
\end{equation}
The NUT$_-$ multiplicities follow analogously after flipping the sign of $n_2$ as expected. Using~\eqref{eq:1-loop-formal} to obtain the formal one-loop determinant, we can compactly write the result for both orientations as
\be
\label{eq:ZNUT-index}
Z^{\mathrm{vec}}_{\textnormal{1-loop}}(\mathrm{NUT}_\pm) = \prod_{n_1, n_2 \geq 1} \Bigl(\frac{ \mathrm{i} n_1 \pm \mathrm{i} \omega\, n_2}{L_b}\Bigr)\,  \prod_{n_1 \geq 1} \Bigl(\frac{ \mathrm{i} n_1}{L_b}\Bigr)^{1/2}\,  \prod_{n_2 \geq 1} \Bigl(\frac{ \pm \mathrm{i} \omega\, n_2}{L_b}\Bigr)^{1/2}\, .
\ee
Once again, we must regularize the infinite products. Taking the logarithm and using zeta-function regularization, we arrive at
\begin{align}
\label{eq:NUTloop}
\log Z^{\mathrm{vec}}_{\textnormal{1-loop}}(\mathrm{NUT}_\pm) =&\; -\sum_{n_1, n_2 \geq 1} \log L_b - \frac12 \sum_{n_1 \geq 1}\log L_b - \frac12  \sum_{n_2 \geq 1} \log L_b + \text{pure number} \nonumber \\
=&\; \frac14\, \log L_b \, ,
\end{align}
where we neglected the pure numbers that do not scale with the length scale~$L_b$. Such terms will contribute a non-trivial function of the equivariant parameter~$\omega$ at order~${\cal O} (N^0)$ in a holographic large~$N$ expansion. By the same argument as the one given below~\eqref{eq:Boltloop}, we can ignore the dependence of~$L_b$ on the particular background and localizing manifold since we are interested in the holographic partition function only at order~$\log N$, and directly relate~$L_b$ to the~$\mathbb{H}_4$ length scale~$L$. Putting together all $(n_V + 1)$ abelian vector multiplets coupled to conformal supergravity, we find the total contribution from a single fixed point to be given by
\begin{equation}
\label{eq:NUTloopfinal}
\log Z^{\mathrm{vec}}_{\textnormal{1-loop}}(\mathrm{NUT}_\pm) = \frac{n_V + 1}4 \log L + \ldots 
\end{equation}
Clearly the information about the orientation of the NUTs and the equivariant parameter~$\omega$ is only subleading, in analogy with the information regarding the orientation of the Bolts. In the language of~\eqref{eq:main}, having in mind that so far we only considered a single fixed point, the above formula shows that each vector multiplet (physical or compensating) contributes~$-1/2$ to the constant~$\alpha$,
\begin{equation}
\label{eq:alphavecNUT}
	\alpha^\mathrm{vec}_\text{NUT} = - \frac12\ .
\end{equation}
This is the same value we found for the contribution of an abelian vector multiplet on a Bolt background, \eqref{eq:alphavecBolt}. We will come back to this in Section~\ref{sec:multiplets}.

%%%%%%%%%%%%%%%%%%%%%%%%%%%%%%%%%%%%%%%%
\subsubsection*{Gluing fixed points}
\label{subsubsec:gluing}
%%%%%%%%%%%%%%%%%%%%%%%%%%%%%%%%%%%%%%%%

Obtaining the contribution of multiplet fixed points for one-loop determinants is straightforward since the Atiyah-Bott formula~\eqref{eq:AB} contains a sum over fixed points. One additional subtlety is that different fixed points can be ``glued'' to each other in different manners, by identifying their respective equivariant parameters $\omega$ according to rules dictated by supersymmetry on different backgrounds. The prime example of this gluing in supergravity was considered in~\cite{Hosseini:2019iad}: in the case of rotating supersymmetric black holes with spherical horizons, there are two fixed points corresponding to the centre of~$\mathbb{H}_2$ and the North and South pole of the~$S^2$. Following this, we now discuss two types of gluing. 

\begin{itemize}
\item  Identity gluing\\
The identity gluing is used for the case of Kerr-Newman black holes, i.e.\ solutions where supersymmetry is preserved without a twist, see \cite{Caldarelli:1998hg,Hristov:2019mqp}. In this case the two fixed points are both of the same orientation, such that $\omega_\text{NP} = \omega_\text{SP}$ as discussed in \cite{Hosseini:2019iad}. This gluing corresponds to the superconformal index (SCI) of the dual field theory, which for the case of a single abelian vector multiplet is then given by
\be
  \log Z^{\mathrm{vec}}_{\textnormal{1-loop}}(\text{SCI}) = 2 \log Z^{\mathrm{vec}}_{\textnormal{1-loop}}(\text{NUT}_+) = \frac12 \log L_b + \ldots
\ee
We find that the detailed information about the gluing is subleading and will not affect the~$\log N$ contribution. 

\item $A$-gluing\\
The $A$-gluing is instead used for the case of rotating twisted black holes, i.e.\ solutions where supersymmetry is preserved with a twist and refinement is added, see \cite{Hristov:2018spe}. In this case the two fixed points are of opposite orientation, such that $\omega_\text{NP} = - \omega_\text{SP}$ as discussed in \cite{Hosseini:2019iad}. This gluing corresponds to the refined topologically twisted index (RTTI) of the dual field theory, and for a single abelian vector multiplet we find
\be
  \log Z^{\mathrm{vec}}_{\textnormal{1-loop}}(\text{RTTI}) = \log Z^{\mathrm{vec}}_{\textnormal{1-loop}}(\text{NUT}_+) + \log Z^{\mathrm{vec}}_{\textnormal{1-loop}}(\text{NUT}_-) =  \frac12 \log L_b + \ldots \, ,
\ee
where again we see that the gluing rules do not affect the $\log N$ contribution.
\end{itemize}

Since the orientation of the NUT is not important for the result~\eqref{eq:NUTloop}, the way we glue various fixed points together does not matter at the level of the one-loop determinants. This will be true not only for the examples discussed above, but in full generality no matter how many fixed points we have and how the gluing rule identifies their respective equivariant parameters~$\omega$. We therefore arrive at the general contribution of $(n_V +1)$ vector multiplets for an arbitrary number $n_{fp}$ of fixed points,
\be
\label{eq:fullNUTl}
\log Z^{\mathrm{vec}}_{\textnormal{1-loop}}(\text{NUT}_\pm) =\frac{n_V + 1}2\, \frac{n_{f}}2\log L + \ldots
\ee
We end by noting that the $A$-gluing example is interesting, since it exhibits a connection between the Bolt and the NUT backgrounds. Indeed, we can take the unrefined limit~$\omega \rightarrow 0$ for the Killing vector~$\xi$ in~\eqref{eq:KVNut}. This limit is smooth and corresponds to having a full~$S^2$ fixed manifold~\cite{Closset:2014pda}, thus reducing to the Bolt case with~$\mathfrak{g} = 0$. Refining by turning on~$\omega$ breaks up the fixed~$S^2$ to two antipodal fixed points, but at order~$\log L$ the result in the limit $\omega \rightarrow 0$ is guaranteed to agree with~\eqref{eq:Boltloopfinal} since~\eqref{eq:fullNUTl} does not depend on~$\omega$.

%%%%%%%%%%%%%%%%%%%%%%%%%%%%%%%%%%%%%%%%
\section{The main formula}
\label{sec:multiplets}
%%%%%%%%%%%%%%%%%%%%%%%%%%%%%%%%%%%%%%%%

%%%%%%%%%%%%%%%%%%%%%%%%%%%%%%%%%%%%%%%%
\subsection{The general argument}
\label{subsec:argument}
%%%%%%%%%%%%%%%%%%%%%%%%%%%%%%%%%%%%%%%%

Following the discussion in the previous section, our general argument explaining why every possible multiplet contributes to the one-loop determinant as in \eqref{eq:main} is now straightforward. We showed how every multiplet contribution can be evaluated via the Atiyah-Singer index theorem, and therefore we expect that the equivariant index of the~$D^{\cal X}_{10}$ differential operator relevant for a given multiplet~$\cal X$ will only depend on the topological data of the background ($\mathfrak{g}_i$ for the Bolts,~$n_f$ for the NUTs) and on the length scale~$L_b$. As discussed, the latter  can be thought of as being equal to the length scale~$L$ of~$\mathbb{H}_4$  at leading order. Furthermore, we saw that additional equivariant parameters such as~$\omega$ and the particular way we decide to glue various fixed points of the canonical isometry~$\xi$ are subleading and do not affect the~$\log L$ coefficient in the free energy.

Thus, for a NUT background with~$n_f$ fixed points, the Atiyah-Bott fixed point theorem shows that the leading contribution of a given multiplet ${\cal X}$ to the one-loop determinant takes the generic form
\be
\label{eq:generalNUT}
	\log Z^{\cal X}_\text{1-loop}(\text{NUT}_\pm) = -\alpha^{\cal X}_\text{NUT}\, \frac{n_{f}}{2} \log L + \ldots \, ,
\ee
where the details of the differential operator~$D_{10}^{\cal X}$ determine the particular value of $\alpha^{\cal X}_{\text{NUT}}$. 

For a Bolt background~$M_4$ with a fixed manifold~$M_q$ isomorphic to~$\Sigma_\mathfrak{g}$ (or a disjoint union thereof), we recall the equivariant index of a generic\footnote{We assume that the operator is elliptic so that the theorem applies, and that the metric of the background is ECC so that the result is free of boundary contributions as discussed previously.} differential operator~$D^{\cal X}_{10}$ as given by the Atiyah-Singer theorem~\cite{shanahan2006atiyah},
\begin{equation}
\label{eq:AS-gen}
\text{ind}_q(D_{10}^{\cal{X}}) = \int_{TM_q} \frac{\mathrm{ch}_q\bigl(j^*\sigma(D_{10}^{\cal X})\bigr)\,\mathrm{Td}(TM_q^\mathbb{C})}{\mathrm{ch}_q\bigl(\bigwedge_{-1} NM_q^\mathbb{C})}\Big\vert_{\text{top}} \, ,
\end{equation}
where~$j\,:\,M_q \longrightarrow M_4$ is the inclusion mapping,~$j^*$ its pullback, and~$\sigma(D)$ denotes the symbol of~$D$ as introduced in Section~\ref{sec:vectors}. Applying this general formula to~$\bar{\partial}_V$ gives back~\eqref{eq:equiv-dolb}. 

The integrand in~\eqref{eq:AS-gen} depends on the details of the operator~$D_{10}^{\cal{X}}$. Without detailed knowledge of this operator, this remains rather abstract. However, this integrand always involves characteristic classes that can be written as polynomials in the Chern classes. Since we are instructed to pick the top-form of such a polynomial and integrate over the (tangent space of the) fixed manifold, and since~$M_q$ is isomorphic to a Riemann surface~$\Sigma_\mathfrak{g}$ or a disjoint union thereof, it follows that the integrand will generically take the form of~$F(\{q\})\,c_1(TM_q)$ for some function~$F$ of the set of equivariant parameters~$\{q\}$. The integration therefore always produces a factor of
\begin{equation}
\int_{TM_q} F(\{q\})\,c_1(TM_q) = 2\,F(\{q\})\,\sum_i (1 - \mathfrak{g}_i) \, ,
\end{equation}
where the dependence on the specific operator~$D_{10}^{\cal X}$ is relegated to the function~$F$. As an example, for the operator relevant for the vector multiplets, the function~$F$ can be read off from~\eqref{eq:equiv-SD-C}. Translating the result for the equivariant index into the one-loop determinant using~\eqref{eq:1-loop-formal} and focusing on the leading contribution to~$\log L$ as before, we therefore conclude that every multiplet~$\mathcal{X}$ will contribute a term
\be
\label{eq:generalBolt}
	\log Z^{\cal X}_\text{1-loop}(\text{Bolt}_\pm) = -\alpha^{\cal X}_\text{Bolt}\, \sum_i (1-\frak{g}_i) \log L + \ldots \, ,
\ee
to the free energy, for some~$L$-independent number~$\alpha^{\cal X}_\text{Bolt}$ derived from the~$F$ function.\\

We can actually say more and give two complimentary arguments to further constrain the form of the~$\log$-corrections. This will show that the coefficients~$\alpha^{\cal X}_\text{NUT}$ and $\alpha^{\cal X}_\text{Bolt}$ entering~\eqref{eq:generalNUT} and~\eqref{eq:generalBolt} must be equal, regardless of the specific form that the relevant differential operators take.

The first argument is based on particular cases of physical interest. As discussed in Section~\ref{subsec:NUTs}, there exists a particular limit in which the NUT and Bolt contributions must agree since one can turn on an omega-deformation and continuously deform the Bolt solution with~$\mathfrak{g} = 0$ to a NUT solution with two fixed points. A concrete example is the case of rotating black holes with a twist. The background has two antipodal fixed points under~$\xi$, and the unrefined limit~$\omega \rightarrow 0$ leads to a blow up of these fixed points into a two-sphere, as explained in~\cite{Benini:2015noa}. Therefore the unrefined limit of a NUT background with~$n_{f} = 2$ must agree with the result for a single Bolt at genus~$\frak{g} = 0$. This imposes 
\be
\label{eq:generalalpha}
\alpha^{\cal X}_\text{NUT} = \alpha^{\cal X}_\text{Bolt} =: \alpha^{\cal X} \quad \forall \; {\cal X} \, .
\ee
This is already manifest in our results for the vector multiplet contribution,~\eqref{eq:alphavecBolt} and~\eqref{eq:alphavecNUT}. The sum of individual contributions~$\alpha^{\cal X}$ for all multiplets in the bulk theory (see also \eqref{eq:sumofalpha} below) will then give the general constant~$\alpha$ in~\eqref{eq:main} and its holographic counterpart in~\eqref{eq:FTmain}.

A second, more mathematical argument that does not rely on the smooth unrefined limit of the omega-deformed backgrounds can also be given. As emphasized many times throughout this paper, the index theorem for abelian gauge groups necessarily relates the~$\log$-corrections to the topological data of the background manifold~$M_4$. As reviewed by Gibbons and Hawking in~\cite{Gibbons:1979xm}, the Euler characteristic of non-compact manifolds with boundary involves a particular combination of~$n_f$ and~$(1 - \mathfrak{g}_i)$, 
\be
	\chi (M_4) = n_f + 2 \sum_i (1 - \mathfrak{g}_i)\, ,
\ee 
We have used this result in our main formula~\eqref{eq:main}, and we can now argue that the contributions from NUTs  and Bolts must indeed come in this particular combination, which again requires that \eqref{eq:generalalpha} holds. Otherwise, one would find an obstruction in rewriting the final result in terms of a the simple topological invariant~$\chi$. There exists of course other topological invariants for a given manifold~$M_4$ (such as the Hirzebruch signature), but in general these do not involve only a simple sum over fixed points and fixed submanifolds under a canonical isometry.

%%%%%%%%%%%%%%%%%%%%%%%%%%%%%%%%%%%%%%%%
\subsection{Hypermultiplets, the gravity multiplet and the massive tower}
\label{subsec:hypers}
%%%%%%%%%%%%%%%%%%%%%%%%%%%%%%%%%%%%%%%%

Having presented the details of the one-loop determinant for vector multiplets in both Bolt and NUT backgrounds and the general argument why analogous expressions hold for arbitrary multiplets, we move to a more explicit discussion for some of the known supergravity multiplets. We will mainly highlight the similarities and important differences compared to vector multiplets. \\

We begin with hypermultiplets. In the superconformal formalism, one of the compensating multiplet required to gauge-fix the theory to the Poincar\'{e} frame can be chosen to be a hypermultiplet. For this particular multiplet, it was shown in~\cite{Hristov:2019xku} that the~$D^\text{hyp}_{10}$ operator that controls the one-loop determinant actually vanishes at the fixed points of the canonical isometry~$\xi$. The compensating hypermultiplet therefore gives a trivial contribution to~$\log Z_{\text{1-loop}}$, regardless of the particular background. In contrast, the equivariant index of the differential operator relevant for a physical hypermultiplet was shown to be equal and opposite to that of a vector multiplet~\cite{Hristov:2019xku}. Following the same logic as for vector multiplets, we therefore arrive at the conclusion that~$n_H$ physical hypermultiplets give a one-loop contribution of
\be
\label{eq:hyp-Bolt}
	\log Z^\mathrm{hyp}_\text{1-loop}(\text{Bolt}_\pm) = -\frac{n_H}{2}\,\sum_i (1 - \frak{g}_i)\log L + \ldots \, ,
\ee
for Bolt backgrounds with disconnected fixed 2-manifolds labeled by~$i$, and of
\be
\label{eq:hyp-NUT}
	\log Z^\mathrm{hyp}_\text{1-loop}(\text{NUT}_\pm) = -\frac{n_H}{2}\,\frac{n_f}{2} \log L + \ldots \, ,
\ee
for NUT backgrounds with~$n_f$ fixed points. In the language of the previous subsection, we thus find
\begin{equation}
\label{eq:alphahyp}
	\alpha^\mathrm{hyp}_\text{NUT} = \alpha^\mathrm{hyp}_\text{Bolt} = \frac12 \, .
\end{equation}
This is another instance where the knowledge of the specific operator~$D_{10}^{\text{hyp}}$ allows for an explicit computation of the~$\alpha^{\cal{X}}$ numbers, and shows that our expectation~\eqref{eq:generalalpha} is indeed borne out of such an analysis. \\

The situation with the Weyl multiplet, giving rise to the massless gravity multiplet on-shell, is considerably more complicated due to the fact that it includes the gauge symmetries of gravity and supersymmetry. In order to properly gauge-fix all the symmetries, one must introduce 51 bosonic ghosts and 54 fermionic ghosts on top of the original 43 bosonic and 40 fermionic degrees of freedom contained in the multiplet~\cite{Jeon:2018kec}. An impressive implementation of the deformed BRST cohomology approach outlined in Section~\ref{sec:setup} was conducted in the latter reference, where the authors found that the cohomological split involves 47 fundamental bosons and 47 fundamental fermions. Based on this, an explicit expression for the relevant differential operator~$D^\text{Weyl}_{10}$ could be given, and its equivariant index computed. However, because the authors of~\cite{Jeon:2018kec} were interested in one-loop determinants relevant for asymptotically flat black holes, they obtained the result in \emph{ungauged} supergravity. There are a number of crucial differences between the localization manifolds and the quadratic fluctuations in gauged and ungauged supergravities, as has been highlighted in~\cite{Hristov:2018lod}. As such, we should be cautious in extrapolating the results of~\cite{Jeon:2018kec} to our current setting. Still, the general arguments of Section~\ref{subsec:argument} can be applied. As a result, for e.g. NUT backgrounds, we expect the Weyl multiplet contribution to be of the form 
\be
\label{eq:Weyl-Bolt}
	\log Z^\mathrm{Weyl}_\text{1-loop}(\text{NUT}_\pm) = -\alpha^\text{Weyl}\,\frac{n_f}{2} \log L + \ldots \, ,
\ee
where~$\alpha^\text{Weyl}$ is a pure number. It is this number that requires a careful analysis of the~$D^\text{Weyl}_{10}$ differential operator.\footnote{For reference, in ungauged supergravity,~\cite{Jeon:2018kec} showed that~$\alpha^\text{Weyl} = 23/6$.} Note that from an on-shell supergravity perspective we recover the one-loop contribution of the Poincar\'{e} gravity multiplet as a combination of the contributions from the Weyl multiplet and the compensating vector multiplet,
\be
	\alpha^\text{grav} = \alpha^\text{Weyl} + \alpha^\text{vec}\ .
\ee

Additional contributions to~$\alpha$ in~\eqref{eq:main} may include other off-shell massless and massive~$\cN=2$ multiplets. Indeed, there exist other matter multiplets one can couple to gauged superconformal gravity, such as tensor, non-linear, or general chiral multiplets. On-shell, these should relate to the allowed short and long massive gravity, gravitino and vector multiplets on AdS$_4$ as detailed in e.g.\ \cite{Klebanov:2008vq}. All these multiplets are known to arise in explicit Kaluza-Klein compactifications from ten and eleven dimensions. The full answer for the coefficient $\alpha$ is therefore given by
\be
\label{eq:sumofalpha}
	\alpha = \alpha^\text{Weyl} + (n_V+1)\, \alpha^\text{vec} + n_H\, \alpha^\text{hyp} + \sum'_\text{KK modes} \alpha^\text{KK}\ ,
\ee
which in principle involves an infinite sum over the KK tower. The latter may need to be appropriately regularized, which we indicated by a prime in the sum.

%%%%%%%%%%%%%%%%%%%%%%%%%%%%%%%%%%%%%
\section{Discussion}
\label{sec:conclusion}
%%%%%%%%%%%%%%%%%%%%%%%%%%%%%%%%%%%%%

In this work we used supergravity localization to study the log-corrections to the on-shell action of asymptotically AdS$_4$ supersymmetric backgrounds, generalizing the results of~\cite{Hristov:2019xku} for static black holes. Our analysis leads to a general formula that factorizes the corrections in terms of a single dynamical coefficient~$\alpha$ and topological data of the background~$M_4$. The formula is given by
\be
\label{eq:maindiscussion}
	- \log Z^\text{sugra}_\text{1-loop} (M_4) = \alpha\,\Bigl(\frac{n_{f}}{2} + \sum_i\,(1 - \frak{g}_i)\Bigr) \log L = \alpha\, \frac{\chi (M_4)}{2}\, \log L\, ,
\ee
which holds at leading order in the holographic scale~$L$ of AdS$_4$. With this result, we conclude with a few additional remarks and open questions.

\begin{itemize}

\item In order to complete our four-dimensional derivation of log-corrections, it is desirable to obtain all explicit contributions to the coefficient~$\alpha$ for all supergravity multiplets, possibly establishing a relation with the conformal anomalies of individual particles as suggested in \cite{Larsen:2015aia}.\footnote{We thank Nikolay Bobev for bringing this point to our attention.} This would then allow for the complete four-dimensional derivation of~$\alpha$, including the entire KK tower. In turn, such a result would find application in many interesting string theory compactifications in which the KK sector is fully understood, see e.g.~\cite{Klebanov:2008vq,Malek:2019eaz,Malek:2020yue,Varela:2020wty} and references therein.

\item A closely related point is that the {\it dynamic} nature of the coefficient~$\alpha$ is tied to the four-dimensional perspective taken in this paper. From a full ten- or eleven-dimensional point of view it is natural to expect that the one-loop result, computed via an index theorem, captures the topological data of both the asymptotically AdS$_4$ space and the internal compact manifold. In turn it would follow that it is precisely the coefficient~$\alpha$ that carries the topological information of the compact space, giving an alternative interpretation of our main formula as a direct factorization of the~$\log$-corrections into two independent topological invariants.\footnote{We thank Alberto Zaffaroni for bringing this point to our attention.} Indeed this is already known to happen for asymptotically flat black holes, where~$\log$-corrections depend on the Euler characteristic of the internal Calabi-Yau three-fold~\cite{Sen:2012kpz}.

\item It is of clear interest to relax the assumptions we have made in our derivation. The possibility of having mildly singular backgrounds~$M_4$ in supergravity like the one of~\cite{Ferrero:2020twa} also prompts us to ask how to apply index theorems on such spaces. Ultimately, we expect~\eqref{eq:maindiscussion} to remain valid for {\it all} supersymmetric backgrounds that asymptote to (Euclidean) AdS$_4$ and it is therefore also desirable to test this result against usual supersymmetric localization results for observables in the 3d boundary theories.

\item As already stated in the introduction, it is of great holographic interest to rewrite~\eqref{eq:maindiscussion} in terms of three-dimensional boundary data, if possible. Such a formulation could greatly facilitate the microscopic understanding of the coefficient~$\alpha$ (or~$\alpha/s$ in the QFT language of~\eqref{eq:FTmain}), and in turn its physical significance. 

\item A specific observation we can make regarding the black holes solutions we have covered is that the fixed-point set of the full spacetime in those examples coincides with the fixed-point set in the near-horizon region. In other words, the log-corrections for black holes can be derived entirely from the knowledge of the near-horizon region. Therefore, we find that the existence of hair degrees of freedom, if any, do not affect the log-corrections studied in this paper.

\item We have focused our attention on holographic partition functions, but of course we are allowed to add various~$Q$-preserving insertions of both local and non-local operators. We could for instance insert Wilson lines in~\eqref{eq:locZ} without introducing major changes in the following localization steps, see e.g.~\cite{Hristov:2018lod}. Thus, we expect our general one-loop formula to hold for various other supersymmetric observables produced by such insertions. However, we need to keep in mind the discussion below~\eqref{eq:logZsaddle}, since additional log-corrections may arise from the saddle-point evaluation of the classical action that will in general also depend on the extra insertions in the path-integral.

\item Put in a more general perspective, our work gives a better low-dimensional understanding of the underlying gravitational structure of $\log$-corrections. Together with the results of~\cite{BenettiGenolini:2019jdz,Hosseini:2019iad} for the classical on-shell action and the results of~\cite{Bobev:2020egg,Bobev:2020zov,Bobev:2021oku} for the perturbative corrections arising from higher-derivative terms, our work makes progress in understanding the bulk structure of the effective four-dimensional theory. It is interesting and desirable to push this program further to higher-order loops by harvesting the full power of exact holography. Supergravity localization seems a promising avenue in this direction, where a natural starting point is to extend the results of~\cite{Dabholkar:2014wpa} and~\cite{Hristov:2018lod,Hristov:2019xku} and generalize them to a broader class of AlAdS solutions. 

\item A natural question is whether one can extend our supergravity analysis to other dimensions. This is indeed easy to imagine given the factorization results for 4d and 5d supersymmetric field theories~\cite{Nekrasov:2002qd,Nekrasov:2003vi}, along with similar gravitational block structure~\cite{Hosseini:2019iad} for different black holes solutions in AdS$_{5,6,7}$. The complete analysis for arbitrary asymptotically AdS solutions in higher dimensions is still lacking. We expect the analog of the results in \cite{BenettiGenolini:2019jdz} in higher dimensions to include also higher dimensional fixed loci such as three- and four-dimensional manifolds. One should then be able to repeat the one-loop analysis here taking into account these generalizations. While the details will differ, we expect that equivariant index theorems will provide the proper mathematical framework also in higher dimensions.

\item An even more ambitious goal is to use our results in non-supersymmetric settings to try and gain a better understanding of $\log$-corrections for thermal black holes. Even if we can no longer rely on the power of localization, it would be interesting if somehow the rich mathematical literature on indices of differential operators could prove relevant to this question.

\end{itemize}

We hope to be able to contribute to these topics in the future.

%%%%%%%%%%%%%%%%%
\section*{Acknowledgements}
We are grateful to Nikolay Bobev, Seyed Morteza Hosseini, and Alberto Zaffaroni for useful discussions, and to Ivano Lodato for collaboration on closely related projects \cite{Hristov:2018lod,Hristov:2019xku}. KH is supported in part by the Bulgarian NSF grants DN08/3, N28/5, and KP-06-N 38/11. VR is supported in part by the KU Leuven C1 grant ZKD1118 C16/16/005.
%%%%%%%%%%%%%%%%%

\begin{appendix}

%%%%%%%%%%%%%%%%%%%%%%%%%%%%%%%%%%%%%%%%
\section{ECC metrics and the fibered Bolt backgrounds}
\label{app:Albin}
%%%%%%%%%%%%%%%%%%%%%%%%%%%%%%%%%%%%%%%%

As explained in Section~\ref{subsec:Bolts}, index theorems for complexes defined on non-compact spaces with a boundary can receive additional contributions compared to their counterpart on closed manifolds. An illustrative example is provided by Albin in~\cite{albin2007renormalized} (Theorem 7.2) for the de Rahm complex, whose index gives the Euler characteristic. The theorem states
\begin{equation}
\label{eq:chi-Albin}
{}^R\int_{M_4} e + \mathrm{FP}_{\varepsilon = 0} \int_{x = \varepsilon} \Theta = \chi(M_4) \, .
\end{equation}
Both integrals above are regularized to deal with the non-compact nature of the manifold~$M_4$. The second term on the left-hand side captures the boundary contribution to~$\chi$, where the boundary~$\partial M_4$ is located at some coordinate~$x = 0$. The regularization introduces a cut-off and extracts the Finite Part (FP) as the cut-off is taken to infinity. Recall that in four dimensions, the Euler class~$e$ and the form~$\Theta$ are given in~\eqref{eq:4deT} in terms of the curvature and second fundamental form. It is further shown in~\cite{albin2007renormalized} that when the metric is Even Conformally Compact (ECC), the boundary contribution to the Euler characteristic vanishes. The definition of an ECC metric is given around~\eqref{eq:CCmetric} in the main text. There, we argued that as long as we use index theorems on manifolds admitting an ECC metric, we can safely ignore the boundary contributions to indices. 

Let us now ask whether the Bolt background falls into the ECC class of metrics. For concreteness, we focus on the case of an~$S^2$ base in what follows, although the general~$\Sigma_\mathfrak{g}$ case follows analogously. The Euclidean Bolt metric is given by~\cite{Toldo:2017qsh}
\begin{equation}
\label{eq:Bolt-metric}
ds^2 = \lambda(r)(d\tau + 2s\cos\theta\,d\phi)^2 + \frac{dr^2}{\lambda(r)} + (r^2 - s^2)\,d\Omega_2^2 \, ,
\end{equation}
in coordinates where the boundary is at~$r\rightarrow + \infty$. The radial function~$\lambda$ is given by
\begin{equation}
\lambda(r) = \frac{(r^2 - s^2)^2 + (1 - 4s^2)(r^2 + s^2) - 2 M r + P^2 - Q^2}{r^2 - s^2} \, ,
\end{equation}
and the solution is supported by a one-form gauge field
\begin{equation}
A = \frac{P(r^2 + s^2) - 2sQ\,r}{r^2 - s^2}\Bigl(\frac{1}{2s}d\tau + \cos\theta\,d\phi\Bigr) \, .
\end{equation}
The parameters~$M$,~$Q$ and~$P$ are related to the mass, the electric charge and the magnetic charge of the solution, while~$s$ characterizes the squashing of the boundary~$S^3$. For the solution to be BPS, we require the parameters to satisfy~\cite{Toldo:2017qsh}
\begin{equation}
P = - \frac12(4s^2 - 1) \, , \qquad M = 2sQ \, .
\end{equation}
We change coordinates as~$r = x^{-1}$ and expand the metric~\eqref{eq:Bolt-metric} near the~$x=0$ boundary to obtain
\begin{equation}
ds^2 \approx \frac{dx^2}{x^2} + \frac{h_{ij}(x)dy^i dy^j}{x^2} \, ,
\end{equation}
where the non-zero elements of the boundary metric are given by
\begin{equation}
\begin{split}
h_{\theta\theta} =&\; 1 - s^2x^2 + \mathcal{O}(x^5) \, , \\
4\,h_{\tau\tau} =&\; 4 + 4(1 - 5s^2)x^2 - 16Qs\,x^3 + (1 - 4Q^2 - 16s^4)\,x^4 + \mathcal{O}(x^5) \, , \\
h_{\tau\phi} =&\; s\cos\theta\,\bigl(4 + 4(1 - 5s^2)x^2 - 16Qs\,x^3 + (1 - 4Q^2 - 16s^4)\,x^4\bigr) + \mathcal{O}(x^5) \, , \\
h_{\phi\phi} =&\; (1 - s^2 x^2)\sin^2\theta \\
&\; + s^2\cos^2\theta\,\bigl(4 + 4(1 - 5s^2)x^2 - 16Qs\,x^3 + (1 - 4Q^2 - 16s^4)\,x^4\bigr) + \mathcal{O}(x^5) \, .
\end{split}
\end{equation}
Note the absence of a linear term in~$x$, and the presence of a cubic term in the expansion proportional to the combination of parameters~$Qs$. This shows that in the general case where~$Qs \neq 0$, the Bolt metric~\eqref{eq:Bolt-metric} is conformally compact but \emph{not} even. Thus, we cannot immediately conclude that the~$\Theta$ contribution in~\eqref{eq:chi-Albin} vanishes. Indeed, a direct computation on this space shows that
\begin{equation}
\mathrm{FP}_{\varepsilon = 0} \int_{x = \varepsilon} \Theta = \frac{2\,Qs}{\pi}\,\beta \, ,
\end{equation}
where $\beta$ is the periodicity of the Euclidean time circle. The failure for the metric to be ECC is controlled by the parameter~$s$ which in turn is related to the Chern degree of the fibration in the topology of the Bolt solution~$\mathcal{O}(-\mathfrak{p}) \rightarrow S^2$. Therefore, a way of ensuring the absence of boundary contributions to the Euler characteristic of a Bolt background is to set~$\mathfrak{p} = 0$ and study the case where the fibration is trivial. This is done in Section~\ref{subsec:Bolts}.

%%%%%%%%%%%%%%%%%%%%%%%%%%%%%%%%%%%%%%%%
\section{Relation to the eleven-dimensional heat kernel approach}
\label{app:11d}
%%%%%%%%%%%%%%%%%%%%%%%%%%%%%%%%%%%%%%%%

For completeness and for the sake of comparison, we give here a very brief summary of the 11d approach to one-loop determinants used previously in~\cite{Bhattacharyya:2012ye,Liu:2017vbl} and references thereof. We will outline the main differences and caveats in the higher-dimensional approach with respect to the one we take in the main part of the paper.

The calculation performed in~\cite{Bhattacharyya:2012ye} showed how to correctly account for the~$\log$ corrections of pure AdS$_4$ with an~$S^3$ boundary by way of heat kernel techniques. This framework (see e.g.~\cite{vassilevich2003heat} for the standard review) is particularly useful in odd dimensions where the Seeley-de Witt coefficients vanish. As such, only the zero-modes of the various fields contribute to the final answer. Those were correctly accounted for in~\cite{Bhattacharyya:2012ye} where the authors considered the 11d space to be a direct product of the maximally symmetric spaces AdS$_4$ and~$S^7$.

An extension of this result to the case of asymptotically AdS$_4$ black holes was presented in~\cite{Liu:2017vbl}, again adopting an eleven-dimensional perspective. However, an important (and as-of-yet unproven) assumption made in~\cite{Liu:2017vbl} is that the 11d space can once again be treated as being split between a four- and a seven-dimensional space, even though asymptotically AdS$_4$ solutions no longer enjoy a direct product structure in eleven dimensions. Under the assumption that the fibration between the asymptotically AdS$_4$ solution and the deformed~$S^7$ does not lead to a different topology,~\cite{Liu:2017vbl} showed that the zero-mode calculation can be reduced to the evaluation of the regularized integral of the Euler class,~${}^R \int e$ on the four-dimensional manifold. Using that the black hole solutions in question do admit ECC metrics, one then obtains the number of zero modes using~\eqref{eq:chi-Albin},
\be
\label{eq:nZM}
	n_0 = {}^R \int_{M_4} e = \chi (M_4) = 2 (1-\mathfrak{g}) \, .
\ee
This result can then be related to one-loop determinant and the~$\log$-correction in the holographically dual field theory prediction, as explained in~\cite{Liu:2017vbl}. From the discussion in Appendix~\ref{app:Albin}, it is clear that~\eqref{eq:nZM} can only be properly understood in the case of ECC metrics, and indeed it is interesting to suggest that the assumption that the fibration does not change the topology might be closely related to the issue of extra boundary modes. 

Let us end this appendix with a short list of pros and cons when comparing our supergravity localization results and the heat kernel method in asymptotically AdS$_4$ spacetimes.

\begin{itemize}

	\item The heat kernel approach is particularly useful for odd-dimensional solutions and is instead rather cumbersome for even-dimensional systems that naturally appear from type II string compactifications. The reason is that non-zero modes entering the general calculation are in principle infinite in AdS due to the lack of scale separation of the KK modes. The similar issue of infinite KK modes complicates also our analysis by preventing us to explicit calculate the coefficient~$\alpha$ in~\eqref{eq:main}. However, our approach makes no particular difference between uplifts to ten and eleven dimensions, thus keeping the outcome general.

	\item The technical assumption that one can further break up the calculation in 4 and 7 dimensions is only correct strictly for pure AdS$_4$ and is otherwise a question that needs to be addressed on a case-by-case basis after a full uplift of the known four-dimensional solutions. In contrast, our approach circumvents this problem as we are able to address the one-loop determinant for general supersymmetric backgrounds. It is however interesting to note that the simplification due to considering ECC metrics seems technically important in both approaches.

	\item The upshot of the direct eleven-dimensional approach is in the explicit evaluation of the dynamical coefficient~$\alpha$ in~\eqref{eq:main} without further assumptions at least in the pure AdS$_4$ case. In this sense one can actually use our general formula~\eqref{eq:FTmain} taking as an extra input the explicit value of~$\alpha$ that, once calculated, holds for the full set of partition functions of the given holographically dual theory.
	
\end{itemize}

These points seem to render the two approaches complementary to each other and likely very useful in combination for deriving~$\log$-corrections in holography.

%%%%%%%%%%%%%%%%%%%%%%%%%%%%%%%%%%%%%%%%
\section{The index of the twisted Dolbeault complex}
\label{app:Shanahan}
%%%%%%%%%%%%%%%%%%%%%%%%%%%%%%%%%%%%%%%%

In this appendix, we give some details on the computation of various characteristic classes entering the computation of the equivariant index of the twisted Dolbeault complex~\eqref{eq:twist-del} via the Atiyah-Singer theorem. We follow the beautiful exposition given in~\cite{shanahan2006atiyah}. Let~$G$ be a compact Lie group acting holomorphically on a compact complex manifold~$M$, let~$M_q \subset M$ be the submanifold of points left fixed by an element~$q \in G$ and let~$V$ be a holomorphic~$G$-bundle on~$M$. The~$G$-equivariant index of the~$V$-twisted Dolbeault operator is given by an integral over the fixed locus,
\begin{equation}
\label{eq:S-Dolb}
\mathrm{ind}_q(\bar{\partial}_{V}) = \int_{M_q}\,\frac{\mathrm{ch}_q(V|M_q)\,\mathrm{Td}(TM^{\,+}_q)}{\mathrm{ch}_q\bigl(1 - NM^{\,-}_q\bigr)}\Big\vert_{\mathrm{top}} \, .
\end{equation} 
As a warm-up, let us consider the case where the bundle~$V$ is trivial and the manifold~$M$ is simply~$\mathbb{R}^4 = \mathbb{C} \times \mathbb{C}$ with complex coordinates~$(z_1,z_2)$ upon which~$G = U(1) \times U(1)$ acts as~$(z_1,z_2) \mapsto (q_1 z_1, q_2 z_2)$. The fixed locus under this action is the origin of $\mathbb{R}^4$, $M_q = \{0\}$. The tangent bundle~$TM_q$ is therefore trivial, and the (complexified) normal bundle is given by~$NM_q = \mathbb{C} \times \mathbb{C}$. The Lie group~$G$ acts holomorphically on the two~$\mathbb{C}$-factors by multiplication by~$q_1$ and~$q_2$. Thus, the~$G$-equivariant Chern character is given by
\begin{equation}
\mathrm{ch}_q\bigl(1 - NM^{\,-}_q\bigr) = (1 - q_1^{-1})(1 - q_2^{-1}) \, . 
\end{equation}
Applying \eqref{eq:S-Dolb}, we see that the integration over~$M_q$ is trivial and that the index is simply
\begin{equation}
\mathrm{ind}_q(\bar{\partial}\,;\,\mathbb{R}^4) = \frac{1}{(1 - q_1^{-1})(1 - q_2^{-1})} \, .
\end{equation}
We can further twist the Dolbeault complex by the bundle $V := \mathcal{O} \oplus K_M$, where~$K_M$ is the canonical bundle over~$M$. The formula above picks up a contribution from the equivariant Chern character of the bundle~$V$ restricted to the fixed locus,
\begin{equation}
\mathrm{ch}_q(V|\{0\}) = 1 + q_1^{-1} q_2^{-1} \, .
\end{equation}
To write the above, we have used the fact that~$K_M = \bigwedge^2 T^*M^+$ is the complex line bundle of holomorphic two-forms, upon which~$G$ acts by multiplication by~$q_1^{-1} q_2^{-1}$. So by \eqref{eq:S-Dolb},
\begin{equation}
\label{eq:twist-dolb}
\mathrm{ind}_q(\bar{\partial}_V\,;\,\mathbb{R}^4) = \frac{1 + q_1 q_2}{(1 - q_1)(1 - q_2)} \, .
\end{equation}
Since the twisted Dolbeault complex is isomorphic to the (complexified) self-dual complex~\eqref{eq:ASD}~\cite{friedman2013smooth}, the above should match the equivariant index of~$D_+$ on~$\mathbb{R}^4$. This is indeed the case, see e.g.~\cite{Pestun:2016jze}. \\

Let us now consider the twisted Dolbeault complex on the 4-sphere $S^4$. In this case, the action of~$G = U(1) \times U(1)$ action has two fixed points, located at the North Pole and at the South Pole. Using~\eqref{eq:twist-dolb}, we obtain the total index
\begin{equation}
\mathrm{ind}_q(\bar{\partial}_V\,;\,S^4) = \left[\frac{1 + q_1 q_2}{(1 - q_1)(1 - q_2)}\right]_{\mathrm{NP}} + \left[\frac{1 + q_1 q_2}{(1 - q_1)(1 - q_2)}\right]_{\mathrm{SP}} \, .
\end{equation}
Although the contributions look equal, the fact that the $\bar{\partial}_V$ operator is not elliptic but only transversally elliptic on the 4-sphere forces us to consider different expansions at the NP and SP. The above agrees with the equivariant index of~$D_+$ as computed in e.g.~\cite{Pestun:2007rz}. For this comparison, it is important to note the overall minus sign coming from folding the SD complex: since $D_+$ is a three-term complex (cf.~\eqref{eq:ASD}), one typically folds it to a standard two-term complex
\begin{equation}
\label{eq:fold}
\delta \oplus \mathrm{d}^+ \; : \; \Omega^1 \longrightarrow \Omega^0 \oplus \Omega^{2+} \, ,
\end{equation}
where $\delta$ is the co-differential. In the folding, the roles of the bosonic and fermionic bundles are switched, and as a result~$\mathrm{ind}_q(D_+) = - \mathrm{ind}_q(\delta \oplus \mathrm{d}^+)$. \\

Moving to the case of interest in Section~\ref{subsec:Bolts}, we now study the twisted Dolbeault complex on~$M = \mathbb{R}^2 \times S^2$ with a $G = U(1)$ group acting as~$(w,z) \mapsto (q w, z)$. The fixed locus is now~$M_q = S^2$. The normal bundle of~$S^2$ in $\mathbb{R}^2 \times S^2$ is trivial,~$NS^2 = \mathbb{R}^2 \times S^2$. The group~$G$ acts on the first factor by multiplication by~$q$, and therefore
\begin{equation}
\label{eq:NMt:BH}
\mathrm{ch}_q\bigl(1 - NM^{\,-}_q\bigr) = 1 - q^{-1} \, .
\end{equation}
Using~$c_1(K_M|S^2) = c_1(K_{S^2}) = -c_1(S^2)$, the equivariant Chern character of the bundle~$V$ is given by
\begin{equation}
\label{eq:VMt:BH}
\mathrm{ch}_q(V|S^2) = 1 + q^{-1}\bigl(1 - c_1(S^2)\bigr) \, .
\end{equation}
Lastly, the Todd class of the tangent bundle is given in terms of the first Chern class of~$M_q = S^2$, 
\begin{equation}
\mathrm{Td}(S^2) = 1 + \frac12\,c_1(S^2) \, .
\end{equation}
Putting these results together in~\eqref{eq:S-Dolb}, we obtain the following equivariant index
\begin{equation}
\begin{split}
\mathrm{ind}_q(\bar{\partial}_V\,;\,\mathbb{R}^2 \times S^2) =&\; \frac{1}{1 - q^{-1}}\,\int_{S^2}\Bigl(1 + \frac12\,c_1(S^2)\Bigr)\Bigl(1 + q^{-1} - q^{-1}\,c_1(S^2)\Bigr)\Big\vert_\mathrm{top} \\[1mm]
=&\; \int_{S^2}\frac12\,c_1(S^2) \, .
\end{split}
\end{equation}
Finally, taking into account the minus sign as explained around~\eqref{eq:fold}, we obtain the equivariant index of the self-dual complex on~$\mathbb{R}^2 \times S^2$,
\begin{equation}
\mathrm{ind}_q(D_+\,;\,\mathbb{R}^2\times S^2) = -\Bigl(\frac{1}{1 - q} + \frac{1}{1 - q^{-1}}\Bigr)\int_{S^2}\frac12\,c_1(S^2) \, .
\end{equation}
Note that the prefactor is formally one, although the way we write it above is convenient to compare to the mode analysis in~\cite{Hristov:2019xku} and the terminology in~\cite{Pestun:2016jze}. The generalization to the case where~$M = \mathbb{R}^2 \times \Sigma_\mathfrak{g}$ is straightforward, and the result is given in~\eqref{eq:equiv-SD-C}.\\

Finally, we comment on the general case relevant for Bolt backgrounds. We consider the twisted Dolbeault complex on~$M \cong \mathcal{O}(-p) \rightarrow \Sigma_\mathfrak{g}$ with a $G = U(1)$ group leaving $M_q = \Sigma_\mathfrak{g}$ fixed. To compute the equivariant Chern character of~$K_M|\Sigma_\mathfrak{g}$ we used the adjunction formula~$K_{M_q} = K_M|M_q \otimes NM_q$ for~$M_q \subset M$ to obtain
\begin{equation}
\mathrm{ch}_q(K_M|\Sigma_\mathfrak{g}) = \frac{q^{-1}\bigl(1 - c_1(\Sigma_\mathfrak{g})\bigr)}{1 + c_1(N\Sigma_\mathfrak{g})} \, ,
\end{equation}
where we used that~$G$ acts on the normal bundle~$N\Sigma_\mathfrak{g}$ by multiplication by~$q$. This leads to the equivariant Chern character of the twisting~$V$-bundle,
\begin{equation}
\mathrm{ch}_q(V|\Sigma_\mathfrak{g}) = 1 + \frac{q^{-1}\bigl(1 - c_1(\Sigma_\mathfrak{g})\bigr)}{1 + c_1(N\Sigma_\mathfrak{g})} \, .
\end{equation}
On the other hand, since the normal bundle~$N\Sigma_\mathfrak{g}$ has non-trivial topology compared to the~$\mathfrak{p} = 0$ case, we also have to take into account its equivariant Chern character. We find
\begin{equation}
\mathrm{ch}_q\bigl(1 -  NM_q{}^-\bigr) = 1 - q^{-1}\bigl(1 - c_1(N\Sigma_\mathfrak{g})\bigr) \, .
\end{equation}
Note that the above two equations simplify to \eqref{eq:VMt:BH} and \eqref{eq:NMt:BH}, respectively, when the normal bundle is trivial. Taking into account the Todd class of $\Sigma_\mathfrak{g}$ and extracting the top-form component, we thus find a contribution to the equivariant index of
\begin{equation}
\label{eq:p-contrib}
\int_{\Sigma_\mathfrak{g}}\,\frac12\,c_1(\Sigma_\mathfrak{g}) - \frac{2q}{(1-q)^2}\,\int_{\Sigma_\mathfrak{g}}\,c_1(N\Sigma_\mathfrak{g}) \, .
\end{equation}
As discussed in the main text and the previous appendix, this may not constitute the full answer for the equivariant index, since there could be non-zero boundary contributions due to the metric not being ECC. Another hint that the above result may be incomplete comes from the fact that the second integral in~\eqref{eq:p-contrib} leads to an explicit~$\mathfrak{p}$-dependence in our formula for the~$\log$-corrections~\eqref{eq:main}. Such a dependence is absent from the results in~\cite{Gang:2019uay}. We hope to be able to better understand and resolve these issues in the future.

\end{appendix}

%%%%%%%%%%%%%%%%%%%%%%%%%%%%%%%%%%%%%
\bibliographystyle{JHEP}
\bibliography{1-loop.bib}

\providecommand{\href}[2]{#2}\begingroup\raggedright\begin{thebibliography}{10}

\bibitem{Pestun:2007rz}
V.~Pestun, {\it {Localization of gauge theory on a four-sphere and
  supersymmetric Wilson loops}},  {\em Commun. Math. Phys.} {\bf 313} (2012)
  71--129, [\href{http://arxiv.org/abs/0712.2824}{{\tt arXiv:0712.2824}}].

\bibitem{Pestun:2016zxk}
V.~Pestun et~al., {\it {Localization techniques in quantum field theories}},
  {\em J. Phys. A} {\bf 50} (2017), no.~44 440301,
  [\href{http://arxiv.org/abs/1608.02952}{{\tt arXiv:1608.02952}}].

\bibitem{Nekrasov:2002qd}
N.~A. Nekrasov, {\it {Seiberg-Witten prepotential from instanton counting}},
  {\em Adv. Theor. Math. Phys.} {\bf 7} (2003), no.~5 831--864,
  [\href{http://arxiv.org/abs/hep-th/0206161}{{\tt hep-th/0206161}}].

\bibitem{Nekrasov:2003vi}
N.~A. Nekrasov, {\it {Localizing gauge theories}},  in {\em {14th International
  Congress on Mathematical Physics}}, 7, 2003.

\bibitem{Beem:2012mb}
C.~Beem, T.~Dimofte, and S.~Pasquetti, {\it {Holomorphic Blocks in Three
  Dimensions}},  {\em JHEP} {\bf 12} (2014) 177,
  [\href{http://arxiv.org/abs/1211.1986}{{\tt arXiv:1211.1986}}].

\bibitem{Closset:2018ghr}
C.~Closset, H.~Kim, and B.~Willett, {\it {Seifert fibering operators in 3d
  $\mathcal{N}=2$ theories}},  {\em JHEP} {\bf 11} (2018) 004,
  [\href{http://arxiv.org/abs/1807.02328}{{\tt arXiv:1807.02328}}].

\bibitem{Dabholkar:2010uh}
A.~Dabholkar, J.~Gomes, and S.~Murthy, {\it {Quantum black holes, localization
  and the topological string}},  {\em JHEP} {\bf 06} (2011) 019,
  [\href{http://arxiv.org/abs/1012.0265}{{\tt arXiv:1012.0265}}].

\bibitem{Dabholkar:2011ec}
A.~Dabholkar, J.~Gomes, and S.~Murthy, {\it {Localization \textbackslash{}\&
  Exact Holography}},  {\em JHEP} {\bf 04} (2013) 062,
  [\href{http://arxiv.org/abs/1111.1161}{{\tt arXiv:1111.1161}}].

\bibitem{Dabholkar:2014wpa}
A.~Dabholkar, N.~Drukker, and J.~Gomes, {\it {Localization in supergravity and
  quantum $AdS_4/CFT_3$ holography}},  {\em JHEP} {\bf 10} (2014) 090,
  [\href{http://arxiv.org/abs/1406.0505}{{\tt arXiv:1406.0505}}].

\bibitem{Murthy:2015yfa}
S.~Murthy and V.~Reys, {\it {Functional determinants, index theorems, and exact
  quantum black hole entropy}},  {\em JHEP} {\bf 12} (2015) 028,
  [\href{http://arxiv.org/abs/1504.01400}{{\tt arXiv:1504.01400}}].

\bibitem{deWit:2018dix}
B.~de~Wit, S.~Murthy, and V.~Reys, {\it {BRST quantization and equivariant
  cohomology: localization with asymptotic boundaries}},  {\em JHEP} {\bf 09}
  (2018) 084, [\href{http://arxiv.org/abs/1806.03690}{{\tt arXiv:1806.03690}}].

\bibitem{BenettiGenolini:2019jdz}
P.~Benetti~Genolini, J.~M. Perez Ipi\~na, and J.~Sparks, {\it {Localization of
  the action in AdS/CFT}},  {\em JHEP} {\bf 10} (2019) 252,
  [\href{http://arxiv.org/abs/1906.11249}{{\tt arXiv:1906.11249}}].

\bibitem{Hosseini:2019iad}
S.~M. Hosseini, K.~Hristov, and A.~Zaffaroni, {\it {Gluing gravitational blocks
  for AdS black holes}},  {\em JHEP} {\bf 12} (2019) 168,
  [\href{http://arxiv.org/abs/1909.10550}{{\tt arXiv:1909.10550}}].

\bibitem{Chamblin:1998pz}
A.~Chamblin, R.~Emparan, C.~V. Johnson, and R.~C. Myers, {\it {Large N phases,
  gravitational instantons and the nuts and bolts of AdS holography}},  {\em
  Phys. Rev. D} {\bf 59} (1999) 064010,
  [\href{http://arxiv.org/abs/hep-th/9808177}{{\tt hep-th/9808177}}].

\bibitem{Martelli:2011fu}
D.~Martelli, A.~Passias, and J.~Sparks, {\it {The gravity dual of
  supersymmetric gauge theories on a squashed three-sphere}},  {\em Nucl. Phys.
  B} {\bf 864} (2012) 840--868, [\href{http://arxiv.org/abs/1110.6400}{{\tt
  arXiv:1110.6400}}].

\bibitem{Martelli:2011fw}
D.~Martelli and J.~Sparks, {\it {The gravity dual of supersymmetric gauge
  theories on a biaxially squashed three-sphere}},  {\em Nucl. Phys. B} {\bf
  866} (2013) 72--85, [\href{http://arxiv.org/abs/1111.6930}{{\tt
  arXiv:1111.6930}}].

\bibitem{Kapustin:2009kz}
A.~Kapustin, B.~Willett, and I.~Yaakov, {\it {Exact Results for Wilson Loops in
  Superconformal Chern-Simons Theories with Matter}},  {\em JHEP} {\bf 03}
  (2010) 089, [\href{http://arxiv.org/abs/0909.4559}{{\tt arXiv:0909.4559}}].

\bibitem{Hama:2011ea}
N.~Hama, K.~Hosomichi, and S.~Lee, {\it {SUSY Gauge Theories on Squashed
  Three-Spheres}},  {\em JHEP} {\bf 05} (2011) 014,
  [\href{http://arxiv.org/abs/1102.4716}{{\tt arXiv:1102.4716}}].

\bibitem{Cacciatori:2009iz}
S.~L. Cacciatori and D.~Klemm, {\it {Supersymmetric AdS(4) black holes and
  attractors}},  {\em JHEP} {\bf 01} (2010) 085,
  [\href{http://arxiv.org/abs/0911.4926}{{\tt arXiv:0911.4926}}].

\bibitem{Katmadas:2014faa}
S.~Katmadas, {\it {Static BPS black holes in U(1) gauged supergravity}},  {\em
  JHEP} {\bf 09} (2014) 027, [\href{http://arxiv.org/abs/1405.4901}{{\tt
  arXiv:1405.4901}}].

\bibitem{Halmagyi:2014qza}
N.~Halmagyi, {\it {Static BPS black holes in AdS$_{4}$ with general dyonic
  charges}},  {\em JHEP} {\bf 03} (2015) 032,
  [\href{http://arxiv.org/abs/1408.2831}{{\tt arXiv:1408.2831}}].

\bibitem{Benini:2015noa}
F.~Benini and A.~Zaffaroni, {\it {A topologically twisted index for
  three-dimensional supersymmetric theories}},  {\em JHEP} {\bf 07} (2015) 127,
  [\href{http://arxiv.org/abs/1504.03698}{{\tt arXiv:1504.03698}}].

\bibitem{Benini:2015eyy}
F.~Benini, K.~Hristov, and A.~Zaffaroni, {\it {Black hole microstates in
  AdS$_{4}$ from supersymmetric localization}},  {\em JHEP} {\bf 05} (2016)
  054, [\href{http://arxiv.org/abs/1511.04085}{{\tt arXiv:1511.04085}}].

\bibitem{Benini:2016hjo}
F.~Benini and A.~Zaffaroni, {\it {Supersymmetric partition functions on Riemann
  surfaces}},  {\em Proc. Symp. Pure Math.} {\bf 96} (2017) 13--46,
  [\href{http://arxiv.org/abs/1605.06120}{{\tt arXiv:1605.06120}}].

\bibitem{Closset:2016arn}
C.~Closset and H.~Kim, {\it {Comments on twisted indices in 3d supersymmetric
  gauge theories}},  {\em JHEP} {\bf 08} (2016) 059,
  [\href{http://arxiv.org/abs/1605.06531}{{\tt arXiv:1605.06531}}].

\bibitem{Benini:2016rke}
F.~Benini, K.~Hristov, and A.~Zaffaroni, {\it {Exact microstate counting for
  dyonic black holes in AdS4}},  {\em Phys. Lett. B} {\bf 771} (2017) 462--466,
  [\href{http://arxiv.org/abs/1608.07294}{{\tt arXiv:1608.07294}}].

\bibitem{Hristov:2018spe}
K.~Hristov, S.~Katmadas, and C.~Toldo, {\it {Rotating attractors and BPS black
  holes in $AdS_4$}},  {\em JHEP} {\bf 01} (2019) 199,
  [\href{http://arxiv.org/abs/1811.00292}{{\tt arXiv:1811.00292}}].

\bibitem{Caldarelli:1998hg}
M.~M. Caldarelli and D.~Klemm, {\it {Supersymmetry of Anti-de Sitter black
  holes}},  {\em Nucl. Phys. B} {\bf 545} (1999) 434--460,
  [\href{http://arxiv.org/abs/hep-th/9808097}{{\tt hep-th/9808097}}].

\bibitem{Hristov:2019mqp}
K.~Hristov, S.~Katmadas, and C.~Toldo, {\it {Matter-coupled supersymmetric
  Kerr-Newman-AdS$_4$ black holes}},  {\em Phys. Rev. D} {\bf 100} (2019),
  no.~6 066016, [\href{http://arxiv.org/abs/1907.05192}{{\tt
  arXiv:1907.05192}}].

\bibitem{Kim:2009wb}
S.~Kim, {\it {The Complete superconformal index for N=6 Chern-Simons theory}},
  {\em Nucl. Phys. B} {\bf 821} (2009) 241--284,
  [\href{http://arxiv.org/abs/0903.4172}{{\tt arXiv:0903.4172}}]. [Erratum:
  Nucl.Phys.B 864, 884 (2012)].

\bibitem{Imamura:2011su}
Y.~Imamura and S.~Yokoyama, {\it {Index for three dimensional superconformal
  field theories with general R-charge assignments}},  {\em JHEP} {\bf 04}
  (2011) 007, [\href{http://arxiv.org/abs/1101.0557}{{\tt arXiv:1101.0557}}].

\bibitem{Closset:2017zgf}
C.~Closset, H.~Kim, and B.~Willett, {\it {Supersymmetric partition functions
  and the three-dimensional A-twist}},  {\em JHEP} {\bf 03} (2017) 074,
  [\href{http://arxiv.org/abs/1701.03171}{{\tt arXiv:1701.03171}}].

\bibitem{Toldo:2017qsh}
C.~Toldo and B.~Willett, {\it {Partition functions on 3d circle bundles and
  their gravity duals}},  {\em JHEP} {\bf 05} (2018) 116,
  [\href{http://arxiv.org/abs/1712.08861}{{\tt arXiv:1712.08861}}].

\bibitem{Atiyah:1974obx}
M.~F. Atiyah, {\em {Elliptic Operators and Compact Groups}}, vol.~401.
\newblock Springer-Verlag, Berline, Germany, 1974.

\bibitem{shanahan2006atiyah}
P.~Shanahan, {\em The Atiyah-Singer index theorem: an introduction}, vol.~638.
\newblock Springer, 2006.

\bibitem{Gibbons:1979xm}
G.~W. Gibbons and S.~W. Hawking, {\it {Classification of Gravitational
  Instanton Symmetries}},  {\em Commun. Math. Phys.} {\bf 66} (1979) 291--310.

\bibitem{Bobev:2020egg}
N.~Bobev, A.~M. Charles, K.~Hristov, and V.~Reys, {\it {The Unreasonable
  Effectiveness of Higher-Derivative Supergravity in AdS$_4$ Holography}},
  {\em Phys. Rev. Lett.} {\bf 125} (2020), no.~13 131601,
  [\href{http://arxiv.org/abs/2006.09390}{{\tt arXiv:2006.09390}}].

\bibitem{Genolini:2021urf}
P.~B. Genolini and P.~Richmond, {\it {The Supersymmetry of Higher-Derivative
  Supergravity in AdS$_4$ Holography}},
  \href{http://arxiv.org/abs/2107.04590}{{\tt arXiv:2107.04590}}.

\bibitem{Bobev:2020zov}
N.~Bobev, A.~M. Charles, D.~Gang, K.~Hristov, and V.~Reys, {\it
  {Higher-derivative supergravity, wrapped M5-branes, and theories of class $
  \mathrm{\mathcal{R}} $}},  {\em JHEP} {\bf 04} (2021) 058,
  [\href{http://arxiv.org/abs/2011.05971}{{\tt arXiv:2011.05971}}].

\bibitem{Bobev:2021oku}
N.~Bobev, A.~M. Charles, K.~Hristov, and V.~Reys, {\it {Higher-Derivative
  Supergravity, AdS$_4$ Holography, and Black Holes}},
  \href{http://arxiv.org/abs/2106.04581}{{\tt arXiv:2106.04581}}.

\bibitem{Banerjee:2010qc}
S.~Banerjee, R.~K. Gupta, and A.~Sen, {\it {Logarithmic Corrections to Extremal
  Black Hole Entropy from Quantum Entropy Function}},  {\em JHEP} {\bf 03}
  (2011) 147, [\href{http://arxiv.org/abs/1005.3044}{{\tt arXiv:1005.3044}}].

\bibitem{Banerjee:2011jp}
S.~Banerjee, R.~K. Gupta, I.~Mandal, and A.~Sen, {\it {Logarithmic Corrections
  to N=4 and N=8 Black Hole Entropy: A One Loop Test of Quantum Gravity}},
  {\em JHEP} {\bf 11} (2011) 143, [\href{http://arxiv.org/abs/1106.0080}{{\tt
  arXiv:1106.0080}}].

\bibitem{Sen:2011ba}
A.~Sen, {\it {Logarithmic Corrections to N=2 Black Hole Entropy: An Infrared
  Window into the Microstates}},  {\em Gen. Rel. Grav.} {\bf 44} (2012), no.~5
  1207--1266, [\href{http://arxiv.org/abs/1108.3842}{{\tt arXiv:1108.3842}}].

\bibitem{Sen:2012cj}
A.~Sen, {\it {Logarithmic Corrections to Rotating Extremal Black Hole Entropy
  in Four and Five Dimensions}},  {\em Gen. Rel. Grav.} {\bf 44} (2012)
  1947--1991, [\href{http://arxiv.org/abs/1109.3706}{{\tt arXiv:1109.3706}}].

\bibitem{Sen:2012dw}
A.~Sen, {\it {Logarithmic Corrections to Schwarzschild and Other Non-extremal
  Black Hole Entropy in Different Dimensions}},  {\em JHEP} {\bf 04} (2013)
  156, [\href{http://arxiv.org/abs/1205.0971}{{\tt arXiv:1205.0971}}].

\bibitem{Charles:2015eha}
A.~M. Charles and F.~Larsen, {\it {Universal corrections to non-extremal black
  hole entropy in $ \mathcal{N}\ge 2 $ supergravity}},  {\em JHEP} {\bf 06}
  (2015) 200, [\href{http://arxiv.org/abs/1505.01156}{{\tt arXiv:1505.01156}}].

\bibitem{Castro:2018hsc}
A.~Castro, V.~Godet, F.~Larsen, and Y.~Zeng, {\it {Logarithmic Corrections to
  Black Hole Entropy: the Non-BPS Branch}},  {\em JHEP} {\bf 05} (2018) 079,
  [\href{http://arxiv.org/abs/1801.01926}{{\tt arXiv:1801.01926}}].

\bibitem{Karan:2021teq}
S.~Karan and B.~Panda, {\it {A generalized Einstein-Maxwell theory:
  Seeley-DeWitt coefficients and logarithmic corrections to the entropy of
  extremal and non-extremal black holes}},
  \href{http://arxiv.org/abs/2104.06381}{{\tt arXiv:2104.06381}}.

\bibitem{Bhattacharyya:2012ye}
S.~Bhattacharyya, A.~Grassi, M.~Marino, and A.~Sen, {\it {A One-Loop Test of
  Quantum Supergravity}},  {\em Class. Quant. Grav.} {\bf 31} (2014) 015012,
  [\href{http://arxiv.org/abs/1210.6057}{{\tt arXiv:1210.6057}}].

\bibitem{Liu:2017vbl}
J.~T. Liu, L.~A. Pando~Zayas, V.~Rathee, and W.~Zhao, {\it {One-Loop Test of
  Quantum Black Holes in anti\textendash{}de Sitter Space}},  {\em Phys. Rev.
  Lett.} {\bf 120} (2018), no.~22 221602,
  [\href{http://arxiv.org/abs/1711.01076}{{\tt arXiv:1711.01076}}].

\bibitem{Gang:2019uay}
D.~Gang, N.~Kim, and L.~A. Pando~Zayas, {\it {Precision Microstate Counting for
  the Entropy of Wrapped M5-branes}},  {\em JHEP} {\bf 03} (2020) 164,
  [\href{http://arxiv.org/abs/1905.01559}{{\tt arXiv:1905.01559}}].

\bibitem{Benini:2019dyp}
F.~Benini, D.~Gang, and L.~A. Pando~Zayas, {\it {Rotating Black Hole Entropy
  from M5 Branes}},  {\em JHEP} {\bf 03} (2020) 057,
  [\href{http://arxiv.org/abs/1909.11612}{{\tt arXiv:1909.11612}}].

\bibitem{Schwarz:2004yj}
J.~H. Schwarz, {\it {Superconformal Chern-Simons theories}},  {\em JHEP} {\bf
  11} (2004) 078, [\href{http://arxiv.org/abs/hep-th/0411077}{{\tt
  hep-th/0411077}}].

\bibitem{Guarino:2015jca}
A.~Guarino, D.~L. Jafferis, and O.~Varela, {\it {String Theory Origin of Dyonic
  N=8 Supergravity and Its Chern-Simons Duals}},  {\em Phys. Rev. Lett.} {\bf
  115} (2015), no.~9 091601, [\href{http://arxiv.org/abs/1504.08009}{{\tt
  arXiv:1504.08009}}].

\bibitem{Liu:2018bac}
J.~T. Liu, L.~A. Pando~Zayas, and S.~Zhou, {\it {Subleading Microstate Counting
  in the Dual to Massive Type IIA}},
  \href{http://arxiv.org/abs/1808.10445}{{\tt arXiv:1808.10445}}.

\bibitem{Liu:2019tuk}
J.~T. Liu and Y.~Lu, {\it {Subleading corrections to the free energy in a
  theory with $N^{5/3}$ scaling}},  {\em JHEP} {\bf 10} (2020) 169,
  [\href{http://arxiv.org/abs/1912.04722}{{\tt arXiv:1912.04722}}].

\bibitem{Marino:2011eh}
M.~Marino and P.~Putrov, {\it {ABJM theory as a Fermi gas}},  {\em J. Stat.
  Mech.} {\bf 1203} (2012) P03001, [\href{http://arxiv.org/abs/1110.4066}{{\tt
  arXiv:1110.4066}}].

\bibitem{Fuji:2011km}
H.~Fuji, S.~Hirano, and S.~Moriyama, {\it {Summing Up All Genus Free Energy of
  ABJM Matrix Model}},  {\em JHEP} {\bf 08} (2011) 001,
  [\href{http://arxiv.org/abs/1106.4631}{{\tt arXiv:1106.4631}}].

\bibitem{Hatsuda:2016uqa}
Y.~Hatsuda, {\it {ABJM on ellipsoid and topological strings}},  {\em JHEP} {\bf
  07} (2016) 026, [\href{http://arxiv.org/abs/1601.02728}{{\tt
  arXiv:1601.02728}}].

\bibitem{Liu:2017vll}
J.~T. Liu, L.~A. Pando~Zayas, V.~Rathee, and W.~Zhao, {\it {Toward Microstate
  Counting Beyond Large N in Localization and the Dual One-loop Quantum
  Supergravity}},  {\em JHEP} {\bf 01} (2018) 026,
  [\href{http://arxiv.org/abs/1707.04197}{{\tt arXiv:1707.04197}}].

\bibitem{Chester:2018aca}
S.~M. Chester, S.~S. Pufu, and X.~Yin, {\it {The M-Theory S-Matrix From ABJM:
  Beyond 11D Supergravity}},  {\em JHEP} {\bf 08} (2018) 115,
  [\href{http://arxiv.org/abs/1804.00949}{{\tt arXiv:1804.00949}}].

\bibitem{PandoZayas:2020iqr}
L.~A. Pando~Zayas and Y.~Xin, {\it {Universal logarithmic behavior in
  microstate counting and the dual one-loop entropy of $AdS_4$ black holes}},
  {\em Phys. Rev. D} {\bf 103} (2021), no.~2 026003,
  [\href{http://arxiv.org/abs/2008.03239}{{\tt arXiv:2008.03239}}].

\bibitem{Gang:2018hjd}
D.~Gang and N.~Kim, {\it {Large $N$ twisted partition functions in 3d-3d
  correspondence and Holography}},  {\em Phys. Rev. D} {\bf 99} (2019), no.~2
  021901, [\href{http://arxiv.org/abs/1808.02797}{{\tt arXiv:1808.02797}}].

\bibitem{Hanada:2012si}
M.~Hanada, M.~Honda, Y.~Honma, J.~Nishimura, S.~Shiba, and Y.~Yoshida, {\it
  {Numerical studies of the ABJM theory for arbitrary N at arbitrary coupling
  constant}},  {\em JHEP} {\bf 05} (2012) 121,
  [\href{http://arxiv.org/abs/1202.5300}{{\tt arXiv:1202.5300}}].

\bibitem{Hong:2021bsb}
J.~Hong and J.~T. Liu, {\it {Subleading corrections to the $S^3$ free energy of
  necklace quiver theories dual to massive IIA}},
  \href{http://arxiv.org/abs/2103.17033}{{\tt arXiv:2103.17033}}.

\bibitem{Aharony:2008ug}
O.~Aharony, O.~Bergman, D.~L. Jafferis, and J.~Maldacena, {\it {N=6
  superconformal Chern-Simons-matter theories, M2-branes and their gravity
  duals}},  {\em JHEP} {\bf 10} (2008) 091,
  [\href{http://arxiv.org/abs/0806.1218}{{\tt arXiv:0806.1218}}].

\bibitem{Imamura:2008nn}
Y.~Imamura and K.~Kimura, {\it {On the moduli space of elliptic
  Maxwell-Chern-Simons theories}},  {\em Prog. Theor. Phys.} {\bf 120} (2008)
  509--523, [\href{http://arxiv.org/abs/0806.3727}{{\tt arXiv:0806.3727}}].

\bibitem{Jafferis:2008qz}
D.~L. Jafferis and A.~Tomasiello, {\it {A Simple class of N=3 gauge/gravity
  duals}},  {\em JHEP} {\bf 10} (2008) 101,
  [\href{http://arxiv.org/abs/0808.0864}{{\tt arXiv:0808.0864}}].

\bibitem{Mezei:2013gqa}
M.~Mezei and S.~S. Pufu, {\it {Three-sphere free energy for classical gauge
  groups}},  {\em JHEP} {\bf 02} (2014) 037,
  [\href{http://arxiv.org/abs/1312.0920}{{\tt arXiv:1312.0920}}].

\bibitem{Hosseini:2016ume}
S.~M. Hosseini and N.~Mekareeya, {\it {Large $N$ topologically twisted index:
  necklace quivers, dualities, and Sasaki-Einstein spaces}},  {\em JHEP} {\bf
  08} (2016) 089, [\href{http://arxiv.org/abs/1604.03397}{{\tt
  arXiv:1604.03397}}].

\bibitem{Emparan:1999pm}
R.~Emparan, C.~V. Johnson, and R.~C. Myers, {\it {Surface terms as counterterms
  in the AdS / CFT correspondence}},  {\em Phys. Rev. D} {\bf 60} (1999)
  104001, [\href{http://arxiv.org/abs/hep-th/9903238}{{\tt hep-th/9903238}}].

\bibitem{Skenderis:2002wp}
K.~Skenderis, {\it {Lecture notes on holographic renormalization}},  {\em
  Class. Quant. Grav.} {\bf 19} (2002) 5849--5876,
  [\href{http://arxiv.org/abs/hep-th/0209067}{{\tt hep-th/0209067}}].

\bibitem{Dunajski:2010uv}
M.~Dunajski, J.~B. Gutowski, W.~A. Sabra, and P.~Tod, {\it {Cosmological
  Einstein-Maxwell Instantons and Euclidean Supersymmetry: Beyond
  Self-Duality}},  {\em JHEP} {\bf 03} (2011) 131,
  [\href{http://arxiv.org/abs/1012.1326}{{\tt arXiv:1012.1326}}].

\bibitem{LopesCardoso:2006ugz}
G.~Lopes~Cardoso, B.~de~Wit, J.~Kappeli, and T.~Mohaupt, {\it {Black hole
  partition functions and duality}},  {\em JHEP} {\bf 03} (2006) 074,
  [\href{http://arxiv.org/abs/hep-th/0601108}{{\tt hep-th/0601108}}].

\bibitem{Denef:2007vg}
F.~Denef and G.~W. Moore, {\it {Split states, entropy enigmas, holes and
  halos}},  {\em JHEP} {\bf 11} (2011) 129,
  [\href{http://arxiv.org/abs/hep-th/0702146}{{\tt hep-th/0702146}}].

\bibitem{Hristov:2019xku}
K.~Hristov, I.~Lodato, and V.~Reys, {\it {One-loop determinants for black holes
  in 4d gauged supergravity}},  {\em JHEP} {\bf 11} (2019) 105,
  [\href{http://arxiv.org/abs/1908.05696}{{\tt arXiv:1908.05696}}].

\bibitem{atiyah1963index}
M.~F. Atiyah and I.~M. Singer, {\it The index of elliptic operators on compact
  manifolds},  {\em Bulletin of the American Mathematical Society} {\bf 69}
  (1963), no.~3 422--433.

\bibitem{Ferrero:2020twa}
P.~Ferrero, J.~P. Gauntlett, J.~M.~P. Ipi\~na, D.~Martelli, and J.~Sparks, {\it
  {Accelerating Black Holes and Spinning Spindles}},
  \href{http://arxiv.org/abs/2012.08530}{{\tt arXiv:2012.08530}}.

\bibitem{Hosseini:2021fge}
S.~M. Hosseini, K.~Hristov, and A.~Zaffaroni, {\it {Rotating multi-charge
  spindles and their microstates}},
  \href{http://arxiv.org/abs/2104.11249}{{\tt arXiv:2104.11249}}.

\bibitem{Cassani:2021dwa}
D.~Cassani, J.~P. Gauntlett, D.~Martelli, and J.~Sparks, {\it {Thermodynamics
  of accelerating and supersymmetric $AdS_4$ black holes}},
  \href{http://arxiv.org/abs/2106.05571}{{\tt arXiv:2106.05571}}.

\bibitem{albin2007renormalized}
P.~Albin, {\it A renormalized index theorem for some complete asymptotically
  regular metrics: the gauss--bonnet theorem},  {\em Advances in Mathematics}
  {\bf 213} (2007), no.~1 1--52.

\bibitem{Lauria:2020rhc}
E.~Lauria and A.~Van~Proeyen, {\em {${\cal N}=2$ Supergravity in $D=4,5,6$
  Dimensions}}, vol.~966.
\newblock Springer, 2020.

\bibitem{deWit:1980lyi}
B.~de~Wit, J.~W. van Holten, and A.~Van~Proeyen, {\it {Structure of N=2
  Supergravity}},  {\em Nucl. Phys. B} {\bf 184} (1981) 77. [Erratum:
  Nucl.Phys.B 222, 516 (1983)].

\bibitem{deWit:2017cle}
B.~de~Wit and V.~Reys, {\it {Euclidean supergravity}},  {\em JHEP} {\bf 12}
  (2017) 011, [\href{http://arxiv.org/abs/1706.04973}{{\tt arXiv:1706.04973}}].

\bibitem{Bobev:2020pjk}
N.~Bobev, A.~M. Charles, and V.~S. Min, {\it {Euclidean Black Saddles and AdS4
  Black Holes}},  \href{http://arxiv.org/abs/2006.01148}{{\tt
  arXiv:2006.01148}}.

\bibitem{Eguchi:1980jx}
T.~Eguchi, P.~B. Gilkey, and A.~J. Hanson, {\it {Gravitation, Gauge Theories
  and Differential Geometry}},  {\em Phys. Rept.} {\bf 66} (1980) 213.

\bibitem{friedman2013smooth}
R.~Friedman and J.~W. Morgan, {\em Smooth four-manifolds and complex surfaces},
  vol.~27.
\newblock Springer Science \& Business Media, 2013.

\bibitem{Hosomichi:2016flq}
K.~Hosomichi, {\it {${{{\mathcal N}}=2}$ SUSY gauge theories on S$^4$}},  {\em
  J. Phys. A} {\bf 50} (2017), no.~44 443010,
  [\href{http://arxiv.org/abs/1608.02962}{{\tt arXiv:1608.02962}}].

\bibitem{Closset:2014pda}
C.~Closset and S.~Cremonesi, {\it {Comments on $ \mathcal{N} $ = (2, 2)
  supersymmetry on two-manifolds}},  {\em JHEP} {\bf 07} (2014) 075,
  [\href{http://arxiv.org/abs/1404.2636}{{\tt arXiv:1404.2636}}].

\bibitem{Jeon:2018kec}
I.~Jeon and S.~Murthy, {\it {Twisting and localization in supergravity:
  equivariant cohomology of BPS black holes}},  {\em JHEP} {\bf 03} (2019) 140,
  [\href{http://arxiv.org/abs/1806.04479}{{\tt arXiv:1806.04479}}].

\bibitem{Hristov:2018lod}
K.~Hristov, I.~Lodato, and V.~Reys, {\it {On the quantum entropy function in 4d
  gauged supergravity}},  {\em JHEP} {\bf 07} (2018) 072,
  [\href{http://arxiv.org/abs/1803.05920}{{\tt arXiv:1803.05920}}].

\bibitem{Klebanov:2008vq}
I.~Klebanov, T.~Klose, and A.~Murugan, {\it {AdS(4)/CFT(3) Squashed, Stretched
  and Warped}},  {\em JHEP} {\bf 03} (2009) 140,
  [\href{http://arxiv.org/abs/0809.3773}{{\tt arXiv:0809.3773}}].

\bibitem{Larsen:2015aia}
F.~Larsen and P.~Lisbao, {\it {Divergences and boundary modes in $
  \mathcal{N}=8 $ supergravity}},  {\em JHEP} {\bf 01} (2016) 024,
  [\href{http://arxiv.org/abs/1508.03413}{{\tt arXiv:1508.03413}}].

\bibitem{Malek:2019eaz}
E.~Malek and H.~Samtleben, {\it {Kaluza-Klein Spectrometry for Supergravity}},
  {\em Phys. Rev. Lett.} {\bf 124} (2020), no.~10 101601,
  [\href{http://arxiv.org/abs/1911.12640}{{\tt arXiv:1911.12640}}].

\bibitem{Malek:2020yue}
E.~Malek and H.~Samtleben, {\it {Kaluza-Klein Spectrometry from Exceptional
  Field Theory}},  {\em Phys. Rev. D} {\bf 102} (2020), no.~10 106016,
  [\href{http://arxiv.org/abs/2009.03347}{{\tt arXiv:2009.03347}}].

\bibitem{Varela:2020wty}
O.~Varela, {\it {Super-Chern-Simons spectra from Exceptional Field Theory}},
  {\em JHEP} {\bf 04} (2021) 283, [\href{http://arxiv.org/abs/2010.09743}{{\tt
  arXiv:2010.09743}}].

\bibitem{Sen:2012kpz}
A.~Sen, {\it {Logarithmic Corrections to N=2 Black Hole Entropy: An Infrared
  Window into the Microstates}},  {\em Gen. Rel. Grav.} {\bf 44} (2012), no.~5
  1207--1266, [\href{http://arxiv.org/abs/1108.3842}{{\tt arXiv:1108.3842}}].

\bibitem{vassilevich2003heat}
D.~V. Vassilevich, {\it Heat kernel expansion: user's manual},  {\em Physics
  reports} {\bf 388} (2003), no.~5-6 279--360.

\bibitem{Pestun:2016jze}
V.~Pestun and M.~Zabzine, {\it {Introduction to localization in quantum field
  theory}},  {\em J. Phys. A} {\bf 50} (2017), no.~44 443001,
  [\href{http://arxiv.org/abs/1608.02953}{{\tt arXiv:1608.02953}}].

\end{thebibliography}\endgroup
%%%%%%%%%%%%%%%%%%%%%%%%%%%%%%%%%%%%%

\end{document}